\documentclass[journal]{IEEEtran}

\usepackage{amsmath}
\usepackage{graphicx}
\usepackage{textcomp}
\usepackage{color}
\usepackage{cases}
\usepackage{cite}
\usepackage{epstopdf}

\ifCLASSINFOpdf

\else

\fi

\begin{document}

\title{Interval Estimation for Conditional Failure Rates of Transmission Lines with Limited Samples}
\author{Ming Yang,~\IEEEmembership{Member,~IEEE,}
	    Jianhui Wang,~\IEEEmembership{Senior Member,~IEEE,}
        Haoran Diao,~\IEEEmembership{Student Member,~IEEE,}
        \\Junjian Qi,~\IEEEmembership{Member,~IEEE,}
        and Xueshan Han% <-this % stops a space
		\thanks{This work is supported in part by the National Science Foundation of China under Grant 51007047 and Grant 51477091, the National Basic Research Program of China (973 Program) under Grant 2013CB228205, and the Fundamental Research Funds of Shandong University. J. Wang's work is supported by the U.S. Department of Energy (DOE)'s Office of Electricity Delivery and Energy Reliability.} 
		\thanks{M. Yang, H. Diao and X. Han are with Key Laboratory of Power System Intelligent Dispatch and Control and Collaborative Innovation Center of Global Energy Internet, Shandong University, Jinan, Shandong 250061 China (e-mail: myang@sdu.edu.cn; shanda10dhr@gmail.com; xshan@sdu.edu.cn).}
		\thanks{J. Wang and J. Qi are with the Energy Systems Division, Argonne National Laboratory, Argonne, IL 60439 USA (e-mail: jianhui.wang@anl.gov; jqi@anl.gov).}
		}

\markboth{Accepted by IEEE Transactions on Smart Grid}{stuff}
\maketitle

\begin{abstract}
The estimation of the conditional failure rate (CFR) of an overhead transmission line (OTL) is essential for power system operational reliability assessment.
It is hard to predict the CFR precisely, although great efforts have been made to improve the estimation accuracy.
One significant difficulty is the lack of available outage samples, due to which the law of large numbers is no longer applicable and no convincing statistical result can be obtained.
To address this problem, in this paper a novel imprecise probabilistic approach is proposed to estimate the CFR of an OTL.
The imprecise Dirichlet model (IDM) is applied to establish the imprecise probabilistic relation between an operational condition and the OTL failure.
Then a credal network is constructed to integrate the IDM estimation results corresponding to various operational conditions and infer the CFR of the OTL.
Instead of providing a single-valued estimation result, the proposed approach predicts the possible interval of the CFR in order to explicitly indicate the uncertainty of the estimation and more objectively represent the available knowledge.
\textcolor{black}{
The proposed approach is illustrated by estimating the CFRs of two LGJ-300 transmission lines located in the same region, and it is also compared with the existing approaches by using data generated from a virtual OTL.
Test results indicate that the proposed approach can obtain much tighter and more reasonable CFR intervals compared with the contrast approaches.}
\end{abstract}


\IEEEpeerreviewmaketitle

\section{Introduction}

\IEEEPARstart{E}{stimation} of the failure rates of overhead transmission lines (OTLs) is crucial for power system reliability assessment, maintenance scheduling and operational risk control \cite{allan2013reliability, billinton1999teaching,qi2013blackout, koval2009assessment,qi2015interaction,qi2016estimating}.
The failure rates of OTLs are usually assumed to be constant and are estimated using the long-term mean values.
However, in reality the failure rates can be significantly influenced by both external and internal operational conditions \cite{williams2003weather}.
\textcolor{black}{
The constant failure rate may work well for the relatively long-term applications, but can lead to erroneous results when applied to the operational risk analysis \cite{ning2011time, feng2008power}.}
The predominant influential factors of the CFR include the aging of the OTL, loading level and external environmental conditions.

Over the last few decades, substantial work has been done on the conditional failure rate (CFR) estimation with respect to various operational conditions.
\textcolor{black}{In \cite{ning2011time}, the time-varying transformer failure probability is investigated, and a delayed semi-Markov process based estimation approach is proposed. }
In \cite{billinton2006application}, the weather conditions are divided into three classes, i.e., normal, adverse and major adverse, to count the failures of different weather conditions.
The influences of multiple weather regions on the failure rate and repair rate of a single transmission line are modeled in \cite{billinton1991novel}.
In \cite{zhou2006modeling}, the Poisson regression and Bayesian network are applied to estimate the conditional failure rates of distribution lines.
Weather-related fuzzy models of failure rate, repair time and unavailability of OTLs are established in \cite{li2008fuzzy}.
The failure rate under both normal and adverse weather conditions is expressed by a fuzzy number in the paper.
It is pointed out in \cite{willis2000panel} that systems with aged components might experience higher than average incidence of failures.
By using the data collected from Electricity of France (EDF), the effects of aging and weather conditions on transmission line failure rates are analyzed in \cite{carer2008weather}, where the influences of wind speed, temperature, humidity, and lightning intensity are examined.
In \cite{bollen2000effects}, the time-varying failure rates are discussed considering the effects of adverse weather and component aging, and data collection efforts for non-constant failure rate estimation are suggested.
%In \cite{christiaanse1971reliability}, the failure rates are modeled for different loading levels, and the occurrence of various loading levels is represented by a two-stage renewal process.

Although great efforts have been made to improve the CFR estimation accuracy, it is still difficult if not impossible to get convincing CFR estimation results.
The major barrier is the lack of historical outage observations under relevant operational conditions.
In this situation, the law of large numbers is no longer applicable and a precise estimation of the failure rate cannot be obtained.
This data-deficient problem might be trivial for the constant failure rate estimation, but is crucial for the condition-related failure rate prediction.

In fact, the uncertainty of failure rate estimation with limited samples has already been recognized.
\textcolor{black}{
In \cite{feng2008power}, an operational risk assessment approach based on the credibility theory is proposed.
In the approach, the uncertainty of the conditional failure probability is modeled by a fuzzy membership function.
Meanwhile, reference \cite{ding2008fuzzy} and \cite{ding2008fuzzy1} build a general fuzzy model to deal with the uncertainty of probabilities and performance levels in the reliability assessment of multi-state systems.
However, the approaches mainly focus on the reliability assessment of the whole system instead of estimating the imprecise failure rate.}

\textcolor{black}{
In \cite{zhou2006modeling}, the Bayesian network is selected as a more preferable approach to model the CFR of overhead distribution lines, and the central limit theorem is adopted to estimate the confidence interval of the predicted CFR.
However, as is well known, the sample size should be large when the theorem is applicable \cite{walpole1993probability}, which may restrict the utilization of the approach in practice.
The central confidence interval of the mean of a Poisson distribution can be equivalently expressed by a Chi-square distribution \cite{johnson2005univariate}.
Therefore, in \cite{li2008fuzzy}, the OTL failure is assumed to follow a Poisson distribution, and then the confidence interval of CFR is estimated according to the relationship between the Chi-square distribution and the Poisson distribution.
The approach avoids using the central limit theorem in the data-insufficiency situation, which increases the practicability of the approach.
However, in the paper only the normal and adverse weather conditions are considered.
Meanwhile, as will be illustrated in Section V, the Chi-square distribution based approach may obtain unreasonable CFR interval when the sample size is small.}

\textcolor{black}{
Imprecise Dirichlet model (IDM) is an efficient approach for the interval-valued probability estimation \cite{walley1991statistical, coolen1997imprecise}.
By using a set of prior probabilities, it can objectively estimate the possible probability interval of a random event. 
In \cite{li2011interval}, based on IDM, an interval-valued reliability analysis approach is proposed for multi-state systems.
Although this approach illustrates the usefulness of the interval-valued failure rate, it ignores the effects of the operational conditions.}
It has to be admitted that the data-insufficiency problem of CFR estimation has not been well addressed.

In this paper, a novel approach based on the imprecise probability theory is proposed to estimate the CFR of an OTL.
%The failure of an OTL is represented by a random variable following the binomial distribution.
Instead of providing a single-valued CFR estimation result, the proposed approach predicts the possible interval of the CFR with limited historical observations, and thus can reflect the available estimation information more objectively.
The IDM is adopted to estimate the imprecise probabilistic relation between the OTL failure and each kind of operational condition.
Then a credal network that can perform imprecise probabilistic reasoning is established to integrate the IDM estimation results corresponding to different operational conditions and obtain the interval-valued CFR.
The advantages of the proposed approach include the following:

\begin{enumerate}
	\item IDM is an efficient statistical approach for drawing out imprecise probabilities from available data.
	By using IDM, the uncertainty of the probabilistic dependency relation between an operational condition and the OTL failure can be properly reflected by the width of the probability interval.
	
	\item The credal network is a flexible probabilistic inference approach. Various operational conditions that have either precise or imprecise probabilistic relations with the OTL failure can be simultaneously considered in the network. Moreover, the occurrence probabilities of the operational conditions can also be represented in the network conveniently.
	
	\item %The proposed approach can truthfully reflect the information contained in the data.
	By combining IDM and the credal network, the proposed approach can explicitly model the uncertainty of the estimated CFR. Meanwhile, as illustrated by the case studies, the CFR interval estimated by the proposed approach is much narrower and more reasonable than that obtained by the contrast approaches.
\end{enumerate}

The rest of this paper is organized as follows:
In Section II a binomial failure rate estimation model is introduced.
Section III discusses the mathematical foundations of the proposed approach.
Details of the CFR estimation are provided in Section IV.
Case studies are presented in Section V
and conclusions are drawn in Section VI.

\section{Model of Failure Rate Estimation}

\subsection{Failure Rate Basics}

Failure rate is a widely applied reliability index that represents the frequency of component failures \cite{finkelstein2008failure}.
Mathematically, the failure rate at time $t$ can be expressed as
\begin{align}\label{eq:fre}
\lambda(t) &= \lim_{\Delta t \to 0} \frac{\Pr \big\{ t < T_f \le t + \Delta t | T_f > t \big\}}{\Delta t},
\end{align}
where
$\lambda (t) $ is the failure rate at time $t$,
$\Pr\{\textrm{A} | \textrm{B} \} $ is the conditional probability of event A given event B,
$T_f$  is the failure time,
$\Delta t$ is a small time interval,
and $\Pr\{t < T_f \le t + \Delta t | T_f > t\}$ is the probability that the OTL fails in the time interval $(t, t + \Delta t]$.

As expressed in Eq. (\ref{eq:fre}), the failure rate indicates an instantaneous probability that a normally operating component breaks down at time $t$.
In practice, the failure rate is usually approximated by the average failure probability within a small time interval $(t, t + \Delta t]$.
This approximation will have sufficient accuracy when $\Delta t$ is so small that the failure rate can be considered as constant within the time interval.
In this paper, hourly OTL failure rates are investigated and thus $\Delta t$ is correspondingly set to 1 hour.
Since the time scale is very small compared with the whole life of OTLs, the hourly failure rate of an OTL can be approximately represented by
\textcolor{black}{
\begin{align}\label{eq:fre1}
\lambda_h(t) \approx \Pr \big\{ t < T_f \le t + 1 | T_f > t \big\},
\end{align}
where $\lambda_h(t)$ is the failure rate in the $t$th hour.}

\vspace*{-6pt}
\subsection{Binomial Model for the Failure Rate Estimation}

\begin{figure}[!htb]
\centering
\includegraphics[width=2.2in]{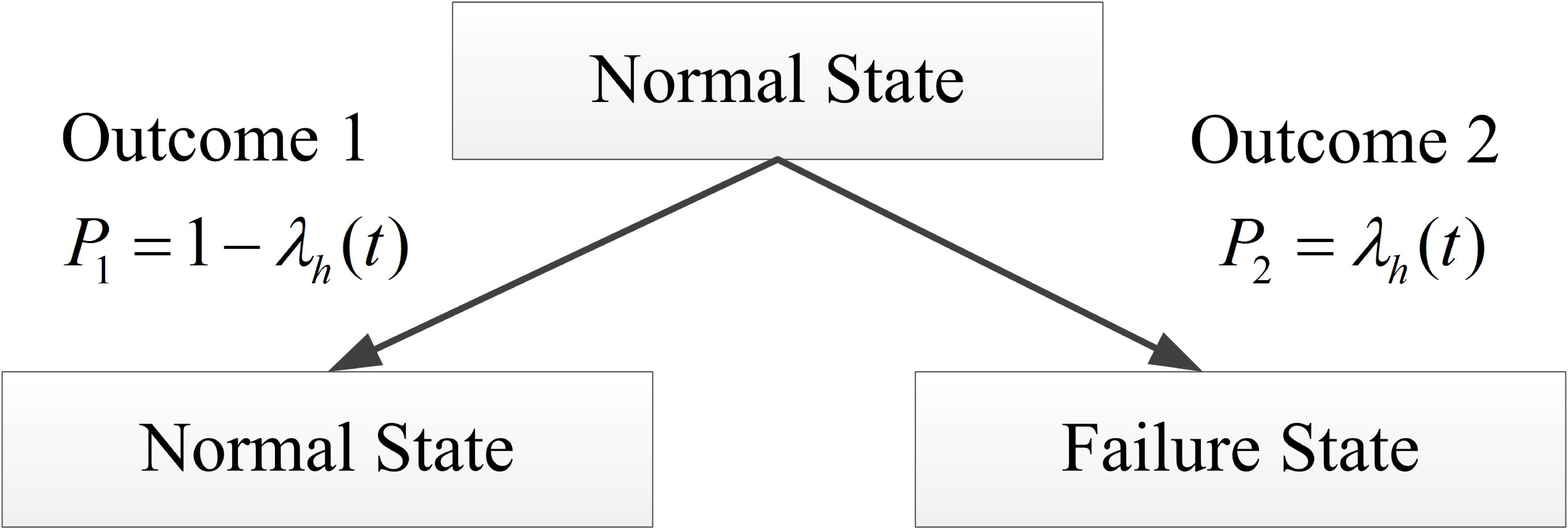}
\caption{Diagram of the binomial distribution model.}
\label{Binomial_Model}
\end{figure}

The hourly failure rate of an OTL can be predicted by estimating the parameter of a binomial distribution.
There are two possible outcomes of the binomial distribution.
One is that the OTL maintains normal operation and the other is that it fails in the relevant hour, as shown in Fig. \ref{Binomial_Model}, where $P_1$ and $P_2$ are two parameters that indicate the possibilities of the outcomes.
According to the definition of the failure rate and aforementioned assumptions, $P_1$ and $P_2$ are equal to $1 - \lambda_h(t)$ and $\lambda_h(t)$, respectively.
So, the hourly failure rate of the OTL can be directly obtained from the estimation result of $P_2$.

This paper will show how to predict the possible interval of $P_2$ under given operational conditions with respect to limited historical observations.
This aim is achieved by using IDM and the credal network, which will be introduced in the following section.
It should be emphasized that the proposed approach is also suitable for the multi-state OTL reliability model.
In that case the multinomial distribution should be applied instead of the binomial distribution used in this paper.

\section{Mathematical Foundations}

\subsection{Imprecise Probability}

Imprecise probability theory is a generalization of classical probability theory allowing partial probability specifications when the available information is insufficient.
The theory bloomed in the 1990s owning to the comprehensive foundations put forward by Walley \cite{walley1991statistical}.

In an imprecise probability model, the uncertainty of each outcome is represented by an interval-valued probability.
For example, the imprecise probabilities of the two outcomes in Fig. \ref{Binomial_Model} can be expressed by $\widetilde{P}_1=[\underline{P}_1,\overline{P}_1]$ and $\widetilde{P}_2=[\underline{P}_2,\overline{P}_2]$  satisfying $ 0 \le \underline{P}_1 \le \overline{P}_1 \le 1$, $0 \le \underline{P}_2 \le \overline{P}_2 \le 1$, $\underline{P}_2=1-\overline{P}_1$, and $\overline{P}_2=1-\underline{P}_1$.

When there is no estimation information at all, the occurrence possibilities of the outcomes will have maximal probability intervals, i.e., $\underline{P}_1=\underline{P}_2=0$  and $\overline{P}_1=\overline{P}_2=1$. If sufficient estimation information is available, the probability interval may shrink to a single point and a precise probability will be obtained \cite{bernard2005introduction}.

\subsection{Imprecise Dirichlet Model}

IDM is an extension of the deterministic Dirichlet model \cite{masegosa2014imprecise}.
Consider a multinomial distribution which has $M$ types of outcomes.
To estimate the occurrence probabilities of the outcomes, the deterministic Dirichlet model uses a Dirichlet distribution as the prior distribution, which is
\begin{align}\label{eq:Dirichlet_prior}
f(P_1,\cdots,P_M)=\frac{\Gamma\!\left(\sum_{m=1}^M a_m\right)}{\prod_{m=1}^M \Gamma(a_m)}\prod_{m=1}^M P_m^{a_m-1},
\end{align}
where
$P_1, P_2, \cdots, P_M$ are the probabilities of the outcomes,
$a_1, a_2, \cdots, a_M$ are the positive parameters of the Dirichlet distribution
and $\Gamma$ is the gamma function.

Since the Dirichlet distribution is a conjugate prior of the multinomial distribution, the posterior of $P_1, P_2, \cdots, P_M$ with respect to the observations also belongs to a Dirichlet distribution, which can be represented as
\begin{align}\label{eq:Dirichlet_posterior}
& f(P_1,\cdots,P_M)
\notag\\
=\ &\frac{\Gamma\!\left(\sum_{m=1}^M a_m+\sum_{m=1}^M n_m\right)}{\prod_{m=1}^M \Gamma(a_m+n_m)}\prod_{m=1}^M P_m^{a_m+n_m-1},
\end{align}
where $n_m$, $m=1,2,\dots,M$, is the number of times that the $m$th outcome is observed.

Therefore, the parameters of the multinomial distribution can be estimated by the expected values of the posterior distribution, as
\begin{align}\label{eq:Dirichlet_result}
P_m = \frac{n_m+a_m}{\sum_{m=1}^M n_m + \sum_{m=1}^M a_m}, \quad  m=1,2,\dots, M.
\end{align}

By analyzing the estimation results of the deterministic Dirichlet model, it can be found that $P_m$ will be $a_m / \sum_{m=1}^M a_m $  if no available observations exist.
This value is the prior estimation of the occurrence possibility of the $m$th outcome.
Here the parameter $a_m$ is called the prior weight of the outcome, and $\sum_{m=1}^M a_m $ is usually denoted by $s$, which is called the equivalent sample size.
In the parameter estimation process, $s$ implies the influence of the prior on the posterior.
The larger $s$ is, the more observations are needed to tune the parameters assigned by the prior distribution.

Equation (\ref{eq:Dirichlet_result}) provides a feasible approach for the multinomial parameter estimation.
However, in practice the available observations may be scarce (like the OTL outage observations).
In this case, the prior weights will significantly influence the parameter estimation results, which causes the results too subjective to be useful.

To avoid the shortcomings of the deterministic Dirichlet model, IDM uses a set of prior density functions instead of a single density function \cite{coolen1997imprecise}, which can be described as
\begin{align}\label{eq:IDM_prior}
f(P_1,\cdots,P_M)= \ &\frac{\Gamma(s)}{\prod_{m=1}^M \Gamma(s\cdot r_m)}\prod_{m=1}^M P_m^{s\cdot r_m-1}, \
\notag\\
&\forall r_m \in [0,1], \  \sum_{m=1}^M r_m=1,
\end{align}
where
$r_m$, $m=1,2,\dots,M$, is the $m$th prior weight factor
and
\textcolor{black}{$s$ is the equivalent sample size.}

In Eq. (\ref{eq:IDM_prior}), $s\cdot r_m$ plays the same role as $a_m$ in Eq. (\ref{eq:Dirichlet_prior}).
When $r_m$ varies in the interval $[0, 1]$, the set will contain all of the possible priors given a predetermined $s$, and the prejudice of the prior can thus be avoided. 

Then the corresponding posterior density function set can be calculated according to Bayesian rules, as
\begin{align}\label{eq:IDM_posterior}
f(P_1,\cdots,P_M)= & \ \frac{\Gamma(s+n)}{\prod_{m=1}^M \Gamma(s\cdot r_m+n_m)}\prod_{m=1}^M P_m^{s\cdot r_m+n_m-1},
\notag\\
& \ \forall r_m \in [0,1],\ \sum_{m=1}^M r_m=1,
\end{align}
where $n=\sum_{m=1}^M n_m$ is the number of total observations.

So, the imprecise parameters of the multinomial distribution can be estimated from the posterior density function set as
\begin{align}\label{eq:IDM_result}
\widetilde{P}_m = [\underline{P}_m,\overline{P}_m]=\left[\frac{n_m}{n+s},\frac{n_m+s}{n+s}\right], \ m=1,2,\dots,M.
\end{align}

The bounds of the probability intervals shown in Eq. (\ref{eq:IDM_result}) are calculated with respect to the bounds of $r_m$, $m=1,2,\dots,M$, and more theoretical details can be found in \cite{walley1991statistical}.

\textcolor{black}{
The interval estimated by IDM  intends to include all the possible probabilities corresponding to different priors.
However, in some applications, the probability interval with a quantitative confidence index may be preferred.
In that case, the probability interval can be estimated by using the credible interval corresponding to the IDM \cite{walley1996inferences}, which is presented in Appendix A.
}

\textcolor{black}{
In fact, several other approaches can also be applied to obtain the probability interval, such as the central limit theorem based approach \cite{zhou2006modeling} and the Chi-square distribution based approach \cite{li2008fuzzy}.
The performance of IDM compared with these existing approaches is illustrated in Appendix B.}

\textcolor{black}{
Moreover, in Eq. (\ref{eq:IDM_result}), the equivalent sample size $s$ is the only exogenous parameter to be specified.
It is set to 1 in our model as suggested by  \cite{bernard2005introduction}.
The effects of $s$ on the IDM estimation result are discussed in Appendix C.
	}

\vspace*{-6pt}
\subsection{Credal Network}

The credal network is developed from the Bayesian network which is a popular graphical model representing the probabilistic relations among the variables of interest \cite{antonucci2007credal, heckerman1998tutorial}.

In a Bayesian network, each node stands for a categorical variable, while the arcs indicate the dependency relations among the variables.
A Bayesian network is composed of a set of conditional mass functions $P(X_i | \boldsymbol{\pi}_i)$, $X_i\in \boldsymbol{X}$ and $\boldsymbol{\pi}_i\in\Omega_{\boldsymbol{\Pi}_i}$,
where $\boldsymbol{X}$ stands for the categorical variables of the network,
$X_i$ is the $i$th categorical variable,
$\boldsymbol{\Pi}_i$ stands for the parent variables of $X_i$,
$\Omega_{\boldsymbol{\Pi}_i}$ is the value space of $\boldsymbol{\Pi}_i$,
and $\boldsymbol{\pi}_i$ is an observation of $\boldsymbol{\Pi}_i$.

\begin{figure}[!b]
	\centering
	\includegraphics[width=3.0in]{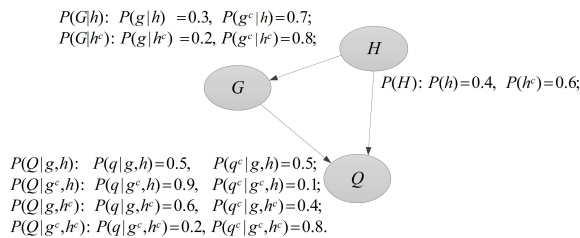}
	\caption{Diagram of a simple Bayesian network.}
	\label{BN}
\end{figure}

A simple Bayesian network is shown in Fig. \ref{BN}, in which the uppercase characters stand for the two-state categorical variables while the lowercase characters stand for the states of the variables.
All necessary conditional mass functions for the inference have been provided in the figure.

A Bayesian network is identical with a joint mass function over $\boldsymbol{X}$ which can be factorized as $P(\boldsymbol{x})=\prod_{i=1}^I P(x_i|\boldsymbol{\pi}_i)$ for each $\boldsymbol{x}\in\Omega_{\boldsymbol{X}}$,
where
$I$ is the number of categorical variables in the network,
$x_i$ is an observation of $X_i$,
$\boldsymbol{x}$ is an observation of $\boldsymbol{X}$,
and $\Omega_{\boldsymbol{X}}$ is the value space of $\boldsymbol{X}$. 
Therefore, with given evidence $\boldsymbol{X}_E=\boldsymbol{x}_E$, the conditional probability of an outcome of the queried variable $X_q$ can be estimated according to Bayesian rules as
\begin{align}\label{eq:BNI}
P(x_q|\boldsymbol{x}_E)=\frac{P(x_q,\boldsymbol{x}_E)}{P(\boldsymbol{x}_E)}=\frac{\sum_{\boldsymbol{X}_M}\prod_{i=1}^I P(x_i|\boldsymbol{\pi}_i)}{\sum_{\boldsymbol{X}_M,X_q}\prod_{i=1}^I P(x_i|\boldsymbol{\pi}_i)},
\end{align}
where
$\boldsymbol{X}_M\equiv\boldsymbol{X}\setminus(\boldsymbol{X}_E\cup X_q)$
and $\sum_{\boldsymbol{X}_M}$ means calculating the total probability with respect to $\boldsymbol{X}_M$ \cite{heckerman1998tutorial}.

For instance, consider the network shown in Fig. \ref{BN}. With evidence $G=g$ and $Q=q$, the conditional probability of $H=h$ can be calculated as
\begin{align*}%\label{eq:ExampleBNI}
& P(h|g,q)=\frac{P(h,g,q)}{P(g,q)}
\notag\\
=\ &\frac{P(h)P(g|h)P(q|g,h)}{P(h)P(g|h)P(q|g,h)+P(h^c)P(g|h^c)P(q|g,h^c)}
\notag\\
=\ &\frac{0.4\times 0.3\times 0.5}{0.4\times 0.3\times 0.5 + 0.6\times 0.2\times 0.6}\approx 0.45.
\end{align*}

The credal network relaxes the Bayesian network by accepting imprecise probabilistic representations \cite{cozman2000credal, cozman2005graphical}.
The differences between the credal network and the Bayesian network lie in the following three levels:

\subsubsection{The probability corresponding to each outcome}

In a Bayesian network, each outcome of a categorical variable has a single-valued probability.
On the contrary, in a credal network, the occurrence chance for each outcome can be represented by an interval-valued probability.   

\subsubsection{The mass function of each categorical variable}
A categorical variable described by imprecise probabilities has a set of mass functions instead of just one.
The convex hull of the mass function set is defined as a credal set \cite{cozman2005graphical, abellan2006measures}, which can be expressed as
\begin{align}\label{eq:CSet}
K(X_i)& =\textrm{CH}\Big\{P(X_i):\text{for each } x_i\in \Omega_{X_i},
\notag\\
P(x_i)& \in\left[\underline{P}(x_i),\overline{P}(x_i)\right]
\text{ and }\sum\nolimits_{\Omega_{X_i}} P(x_i)=1 \Big\},
\end{align}
where
$K(X_i)$ is the credal set of $X_i$,
$\textrm{CH}$ means a convex hull,
$\{P(X_i)$: $\text{Descriptions}\}$ is a set of probability mass functions specified by the descriptions,
$\Omega_{X_i}$ is the value space of $X_i$,
and $\sum_{\Omega_{X_i}} P(x_i) = 1$ means the sum of the probabilities should be equal to 1.

Although the credal set contains an infinite number of mass functions, it only has a finite number of extreme mass functions, which are denoted by $\textrm{EXT}[K(X_i)]$.
These extreme mass functions correspond to the vertexes of the convex hull and can be obtained by combining the endpoints of the probability intervals.
For instance, assume $P(h)$ and $P(h^c)$ in Fig. \ref{BN} are imprecise, i.e., $P(h)\in[0.3,0.5]$ and $P(h^c)\in[0.5,0.7]$.
In this case, the mass function $P(H)$ can be any combination of $P(h)$ and $P(h^c)$ satisfying the constraint $P(h)+P(h^c)=1$.
However, $P(H)$ only has two extreme mass functions, which are $\{P(h) = 0.3, P(h^c) = 0.7\}$ and $\{P(h) = 0.5, P(h^c) = 0.5\}$.

The inference over a credal network is based on the conditional credal sets $K(X_i|\boldsymbol{\pi}_i)$, $i=1,2,\dots,I$.
And it is equivalent to the inference based only on the extreme mass functions of the conditional credal sets.

\subsubsection{The joint mass function of the network}
While a Bayesian network defines a joint mass function, a credal network defines a joint credal set that contains a collection of joint mass functions. 
The convex hull of these joint mass functions is called a strong extension of the credal network \cite{corani2012bayesian}, and can be formulated as
\begin{align}\label{eq:StrongExtension}
K(\boldsymbol{X})& = \textrm{CH}\Big\{P(\boldsymbol{X}):\text{for each } \boldsymbol{x}\in\Omega_{\boldsymbol{X}},
\notag\\
P(\boldsymbol{x})& =\prod_{i=1}^I P(x_i|\boldsymbol{\pi}_i), \ P(X_i|\boldsymbol{\pi}_i)\in K(X_i|\boldsymbol{\pi}_i)\Big\},
\end{align}
where
$K(\boldsymbol{X})$ is the strong extension of the credal network,
$\{P(\boldsymbol{X})$: $\text{Descriptions}\}$ is a set of joint mass functions specified by the descriptions,
and $P(X_i|\boldsymbol{\pi}_i)\in K(X_i|\boldsymbol{\pi}_i)$ indicates that the conditional mass function $P(X_i|\boldsymbol{\pi}_i)$ should be selected from the conditional credal set $K(X_i|\boldsymbol{\pi}_i)$.

Observe from Eq. (\ref{eq:StrongExtension}) that the joint probability $P(\boldsymbol{x})$ is imprecise since each $P(x_i|\boldsymbol{\pi}_i)$, whose mass function $P(X_i|\boldsymbol{\pi}_i)$ can be arbitrarily selected from the credal set $K(X_i|\boldsymbol{\pi}_i)$, is imprecise.
For this reason, $K(\boldsymbol{X})$ contains a set of joint mass functions.
On the other hand, when the mass functions $P(X_i|\boldsymbol{\pi}_i)$, $i=1,2,\dots,I$, are selected, the joint probability $P(\boldsymbol{x})$ for each $\boldsymbol{x}\in\Omega_{\boldsymbol{X}}$ will be determined, and hence one deterministic joint mass function $P(\boldsymbol{X})$ will be obtained.
It is easy to imagine that each joint mass function of a strong extension corresponds to a Bayesian network.

If $P(X_i|\boldsymbol{\pi}_i)$ can only be selected from the extreme mass functions of $K(X_i|\boldsymbol{\pi}_i)$, the resulted joint mass functions are called the extreme joint mass functions, and they correspond to the vertexes of the strong extension.
The extreme joint mass functions can be obtained by
\begin{align}\label{eq:ExtremeJointMassFunction}
& \textrm{EXT}[K(\boldsymbol{X})] = \Big\{P(\boldsymbol{X}):\text{for each } \boldsymbol{x}\in\Omega_{\boldsymbol{X}},
\notag\\
& P(\boldsymbol{x})=\prod_{i=1}^I P(x_i|\boldsymbol{\pi}_i),P(X_i|\boldsymbol{\pi}_i)\in \textrm{EXT}[K(X_i|\boldsymbol{\pi}_i)]\Big\}.
\end{align}

Inference over a credal network is to compute the probability bounds of $X_q=x_q$ respecting $\boldsymbol{X}_E=\boldsymbol{x}_E$.
According to the association between the credal network and the Bayesian network, this aim can be achieved by inferring over the Bayesian networks corresponding to the extreme joint mass functions, as
\begin{align}\label{eq:CNInference}
& \overline{P}(X_q=x_q|\boldsymbol{X}_E=\boldsymbol{x}_E)
\notag\\
=\ &\max\limits_{P(\boldsymbol{X})\in\textrm{EXT}[K(\boldsymbol{X})]}\frac{P(X_q=x_q,\boldsymbol{X}_E=\boldsymbol{x}_E)}{P(\boldsymbol{X}_E=\boldsymbol{x}_E)},
\notag\\[0.2cm]
& \underline{P}(X_q=x_q|\boldsymbol{X}_E=\boldsymbol{x}_E)
\notag\\
=\ &\min\limits_{P(\boldsymbol{X})\in\textrm{EXT}[K(\boldsymbol{X})]}\frac{P(X_q=x_q,\boldsymbol{X}_E=\boldsymbol{x}_E)}{P(\boldsymbol{X}_E=\boldsymbol{x}_E)},
\end{align}
where
$P(\boldsymbol{X})\in\textrm{EXT}[K(\boldsymbol{X})]$ indicates that $P(\boldsymbol{X})$ should be selected from the extreme joint mass functions of $K(\boldsymbol{X})$.

It can be seen from Eq. (\ref{eq:CNInference}) that when an extreme joint mass function is selected, Eq. (\ref{eq:CNInference}) will become a Bayesian network inference problem which can be conveniently solved by Eq. (\ref{eq:BNI}).
Therefore, inference over a credal network can be executed by performing the following four steps:

\begin{enumerate}
\item For each categorical variable $X_i$ within the network, calculate its extreme conditional mass function $\textrm{EXT}[K(X_i|\boldsymbol{\pi}_i)]$ by combining the probability interval endpoint $\underline{P}(x_i|\boldsymbol{\pi}_i)$ and $\overline{P}(x_i|\boldsymbol{\pi}_i)$ of all $x_i\in\Omega_{X_i}$.  
\item Form the extreme joint mass functions $\textrm{EXT}[K(\boldsymbol{X})]$ with respect to the extreme conditional mass functions $\textrm{EXT}[K(X_i|\boldsymbol{\pi}_i)]$, $i=1,2,\dots,I$, $\boldsymbol{\pi}_i\in\Omega_{\boldsymbol{\Pi}_i}$, according to Eq. (\ref{eq:ExtremeJointMassFunction}).
\item Infer on each extreme joint mass function of $\textrm{EXT}[K(\boldsymbol{X})]$ with given evidence $\boldsymbol{X}_E=\boldsymbol{x}_E$, and obtain the conditional probability $P(X_q=x_q|\boldsymbol{X}_E=\boldsymbol{x}_E)$ according to Eq. (\ref{eq:BNI}).
\item Find the maximum and minimum probabilities of $P(X_q=x_q|\boldsymbol{X}_E=\boldsymbol{x}_E)$ with respect to the inference results of all the extreme joint mass functions.
\end{enumerate}

\section{Details of Imprecise CFR Estimation}

\subsection{Imprecise CFR Estimation Model}

\begin{figure}[!htb]
	\centering
	\includegraphics[width=3.0in]{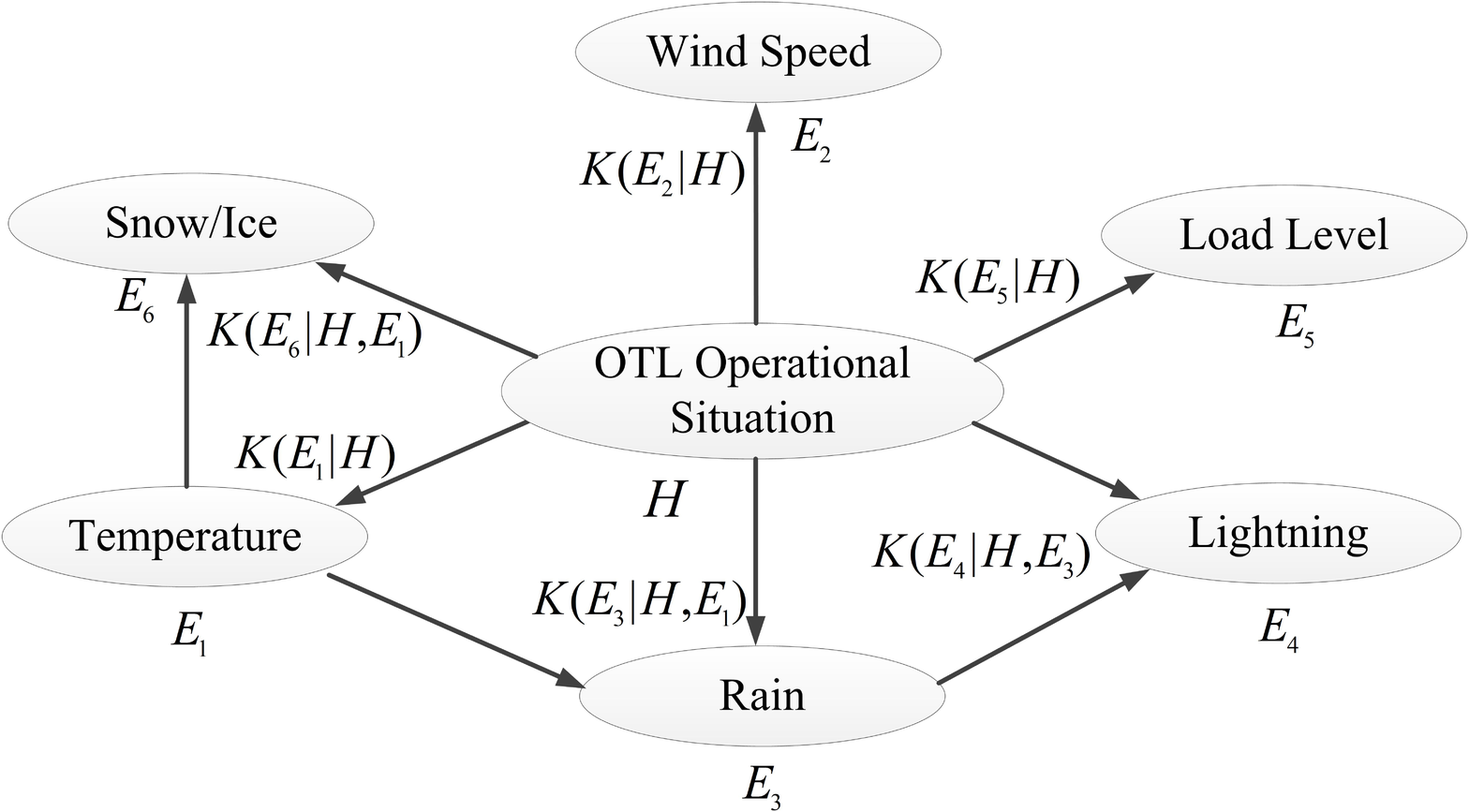}
	\caption{\textcolor{black}{Credal network for the imprecise CFR estimation.}}
	\label{CNforCFR}
\end{figure}

The failure rate of an OTL has a close relationship with surrounding weather conditions.
An OTL is more likely to break down in adverse weather conditions, e.g., \textcolor{black}{high air temperature, strong wind, snow/ice, rain, and lightning}.
%\textcolor{black}{Especially, the accumulated snow and icing can pose dramatic load and tension on an OTL, which may increase the failure rate of the line.}
The loading condition can also influence the failure rate of an OTL. When carrying heavy load, an OTL may suffer from a higher probability of a short-circuit fault \cite{christiaanse1971reliability}.
Moreover, the failure rate of an OTL may change gradually with the age of the line.
An aged transmission line usually has a relatively higher failure rate under the same weather and loading conditions.
Taking these factors into account, a credal network is established for estimating the CFR of an OTL, as shown in Fig. \ref{CNforCFR}.

In the figure, the operational conditions are denoted by the categorical variables $E_1, E_2,\cdots, E_6$, and the state of the OTL is indicated by the binary variable $H$.
Meanwhile, the dependency relations between the OTL state and the operational conditions are reflected by the conditional credal set $K(E_1|H)$, $K(E_2|H)$, $K(E_3 | H, E_1)$, $K(E_4 | H, E_3)$, $K(E_5|H)$ and $K(E_6 | H, E_1)$.
It should be emphasized that the dependency relations among the operational conditions are also reflected in the network.

\textcolor{black}{
The aging of the OTL is not explicitly included in the network. Instead, the average failure rate of similar age OTLs is integrated into the inference model as a prior failure rate,
by which the aging effects on the failure rate is respected.
}

The states of the categorical variables are listed in Table \ref{Table_StatesofCategoricalVariable}, where $T$ is the hourly average temperature, $W$ is the hourly average wind speed, and $L$ is the hourly average loading rate.
\textcolor{black}{
Theoretically, the condition values having similar influences on the failure rate should be classified into the same state to reduce the loss of information \cite{zhou2006modeling}.
However, due to the restriction of data, the states listed in Table \ref{Table_StatesofCategoricalVariable} are categorized according to the classification criteria of utilities' outage records.}

%Moreover, the classification should be different for distinct applications. 

\begin{table}[!htb]
\renewcommand{\arraystretch}{1.2}
\caption{\textcolor{black}{States of the Categorical Variables}}
\vspace*{-6pt}
\label{Table_StatesofCategoricalVariable}
\newcommand{\tabincell}[2]{\begin{tabular}{@{}#1@{}}#2\end{tabular}}
\centering
\begin{tabular}{c c c c}
\hline
Variables & State 1 & State 2 & State 3\\
\hline
$E_1$ & \tabincell{c}{$T\le$4\textcelsius\\($e_{1,1}$)} & \tabincell{c}{4\textcelsius$<T\le$26\textcelsius\\($e_{1,2}$)} & \tabincell{c}{$T>$26\textcelsius\\($e_{1,3}$)}\\
$E_2$ & \tabincell{c}{$W\le$12km/h\\($e_{2,1}$)}        & \tabincell{c}{12km/h$<W\le$40km/h\\($e_{2,2}$)}                & \tabincell{c}{$T>$40km/h\\($e_{2,3}$)}        \\
$E_3$ & \tabincell{c}{Rain\\($e_{3,1}$)}                & \tabincell{c}{No Rain\\($e_{3,2}$)}                            &  N/A                                          \\
$E_4$ & \tabincell{c}{Lightning\\($e_{4,1}$)}           & \tabincell{c}{No Lightning\\($e_{4,2}$)}                       &  N/A                                          \\
$E_5$ & \tabincell{c}{$L\le80\%$\\($e_{5,1}$)}          & \tabincell{c}{$L>80\%$\\($e_{5,2}$)}                           &  N/A                                          \\
\textcolor{black}{$E_6$} & \tabincell{c}{\textcolor{black}{Snow/Ice} \\ \textcolor{black}{($e_{6,1}$)}}          & \tabincell{c}{\textcolor{black}{No Snow/Ice}\\ \textcolor{black}{($e_{6,2}$)}}                          &  \textcolor{black}{N/A}                                          \\
$H$   & \tabincell{c}{Normal Operation\\($h_1$)}    & \tabincell{c}{Failure\\($h_2$)}                                    &  N/A                                          \\
\hline
\end{tabular}
\end{table}

%It can be observed from the table that the continuous temperature, wind speed and loading rate are classified into discrete states for the application of the proposed approach. Theoretically, it is supposed that the operational conditions having similar influence on the failure rate are classified into the same state \cite{zhou2006modeling}. However, since only the categorized condition data can be collected from utility's outage records, TABLE I just lists the state classification criteria of the records.

\vspace*{-10pt}
\subsection{Estimation of the Extreme Mass Functions}
Estimating the extreme mass functions of the credal sets $K(E_1|H)$, $K(E_2|H)$, $K(E_3 | H, E_1)$, $K(E_4 | H, E_3)$, $K(E_5|H)$ and $K(E_6 | H, E_1)$ is the first step for the CFR inference.
To achieve this purpose, the endpoints of the conditional probabilities corresponding to $P(E_1|H)$, $P(E_2|H)$, $P(E_3 | H, E_1)$, $P(E_4 | H, E_3)$, $P(E_5|H)$ and $P(E_6 | H, E_1)$ have to be calculated using IDM, as expressed in Eq. (\ref{eq:IDM_result}).

For instance, to estimate the extreme mass functions of the conditional credal set $K(E_5 | h_2)$, which describes the uncertainty of the loading condition when the transmission line fails, the endpoints of the probability intervals corresponding to $P(e_{5,1}|h_2)$ and $P(e_{5,2}|h_2)$ have to be calculated, as
\begin{align*}%\label{eq:endpointsexample}
& \underline{P}(e_{5,1}|h_2)=\frac{n_{NL}}{n_{F}+s}, \ \ \overline{P}(e_{5,1}|h_2)=\frac{n_{NL}+s}{n_{F}+s},
\notag\\
& \underline{P}(e_{5,2}|h_2)=\frac{n_{HL}}{n_{F}+s}, \ \ \overline{P}(e_{5,2}|h_2)=\frac{n_{HL}+s}{n_{F}+s},
\end{align*}
where 
$n_{NL}$ and $n_{HL}$ are the numbers of samples for which the OTL breaks down with a normal and high loading rate, respectively,
and $n_{F} = n_{NL} + n_{HL}$ is the number of samples for which the OTL breaks down in the relevant time interval.

Then the conditional credal set $K(E_5 | h_2)$ can be obtained from Eq. (\ref{eq:CSet}) as
\begin{align*}%\label{eq:credalsetexample}
K(E_5|h_2) & =\textrm{CH}\Big\{P(E_5|h_2):
\notag\\
& P(e_{5,1}|h_2)\in[\underline{P}(e_{5,1}|h_2),\overline{P}(e_{5,1}|h_2)],
\notag\\
& P(e_{5,2}|h_2)\in[\underline{P}(e_{5,2}|h_2),\overline{P}(e_{5,2}|h_2)],
\notag\\
& P(e_{5,1}|h_2)+P(e_{5,2}|h_2)=1\Big\}.
\end{align*}

Also, the extreme mass functions of the credal set can be obtained by combining the endpoints of the probability intervals as 
\begin{align*}%\label{eq:extrememassfunctionexample}
\textrm{EXT} & [K(E_5|h_2)]=
\notag\\
& \Big\{P(e_{5,1}|h_2)=\underline{P}(e_{5,1}|h_2),P(e_{5,2}|h_2)=\overline{P}(e_{5,2}|h_2)\Big\}
\notag\\
\text{and}  \ & \Big\{P(e_{5,1}|h_2)=\overline{P}(e_{5,1}|h_2),P(e_{5,2}|h_2)=\underline{P}(e_{5,2}|h_2)\Big\}.
\end{align*}

Other conditional credal sets and corresponding extreme mass functions can be calculated in the same way.

\subsection{Imprecise CFR Inference with the Credal Network}

With respect to the credal network mentioned above, the bounds of the CFR under specified operational conditions can be calculated by the equations deduced from Eqs. (\ref{eq:BNI}), (\ref{eq:ExtremeJointMassFunction}) and (\ref{eq:CNInference}), as
\begin{align}\label{eq:CFRInference}
&\overline{\lambda}=\overline{P}(h_2|\boldsymbol{e})
=\max\dfrac{\prod_{j=1}^6\mathcal{P}_{2,j}\cdot P(h_2)}{\sum_{i=1}^2\prod_{j=1}^6\mathcal{P}_{i,j}\cdot P(h_i)},
\notag\\
&\underline{\lambda}=\underline{P}(h_2|\boldsymbol{e})
=\min\dfrac{\prod_{j=1}^6\mathcal{P}_{2,j}\cdot P(h_2)}{\sum_{i=1}^2\prod_{j=1}^6\mathcal{P}_{i,j}\cdot P(h_i)},
\notag\\
& \text{when} \ \ j=1,2,5,
\notag\\
&\mathcal{P}_{2,j} = P(e_{j,k_j}|h_2),\ \ \ \ \ \ \ \ \mathcal{P}_{i,j} = P(e_{j,k_j}|h_i),
\notag\\
& \text{when} \ \ j=3,
\notag\\
&\mathcal{P}_{2,j} = P(e_{3,k_3}|h_2,e_{1,k_1}),\ \mathcal{P}_{i,j} = P(e_{3,k_3}|h_i,e_{1,k_1}),
\notag\\
& \text{when} \ \ j=4,
\notag\\
&\mathcal{P}_{2,j} = P(e_{4,k_4}|h_2,e_{3,k_3}),\ \mathcal{P}_{i,j} = P(e_{4,k_4}|h_i,e_{3,k_3}),
\notag\\
& \text{when} \ \ j=6,
\notag\\
&\mathcal{P}_{2,j} = P(e_{6,k_6}|h_2,e_{1,k_1}),\ \mathcal{P}_{i,j} = P(e_{6,k_6}|h_i,e_{1,k_1}),
%\notag\\
%& \ \ \ \ P(E_j|h_i) \in \textrm{EXT}[K(E_j|h_i)],
%\notag\\
%& \ \ \ \ P(E_4|h_i,e_{3,k_3}) \in \textrm{EXT}[K(E_4|h_i,e_{3,k_3})];
%\notag\\
%& \ \ \ \ P(E_6|h_i,e_{1,k_1}) \in \textrm{EXT}[K(E_6|h_i,e_{1,k_1})];
%\notag\\
%& \ \ \ \ i=1,2;\ \ j=1,2,3,4,5,6;
%\notag\\
%& \ \ \ \ \text{when} \ \ j=1,2, \ \ k_j=1,2,3;
%\notag\\
%& \ \ \ \ \text{when} \ \ j=3,4,5,6 \ \ k_j=1,2
\end{align}
where
$i$ is the index of OTL's states,
$j$ is the index of the evidence variables,
$k_j$ is used to specify the state of the $j$th evidence variable,
\textcolor{black}{$P(h_2)$ is the prior probability of $h_2$ which is assigned by using the average failure rate of the OTLs within the same age group,}
$P(h_1)$ $=$ $1-P(h_2)$, 
$P(e_{j,k_j} | h_i)$ is the conditional probability of the observation $e_{j, k_j}$ given $h_i$ with respect to the conditional mass function $P(E_j | h_i)$,
$P(e_{3,k_3} | h_i, e_{1,k_1})$ is the conditional probability of $e_{3,k_3}$ given $h_i$ and $e_{1,k_1}$ with respect to $P(E_3 | h_i, e_{1,k_1})$,
$P(e_{4,k_4} | h_i, e_{3,k_3})$ is the conditional probability of $e_{4,k_4}$ given $h_i$ and $e_{3,k_3}$ with respect to $P(E_4 | h_i, e_{3,k_3})$,
and 
$P(e_{6,k_6} | h_i, e_{1,k_1})$ is the conditional probability of $e_{6,k_6}$ given $h_i$ and $e_{1,k_1}$ with respect to $P(E_6 | h_i, e_{1,k_1})$.
Meanwhile, the conditional mass function $P(E_j | h_i)$, $P(E_3 | h_i, e_{1,k_1})$, $P(E_4 | h_i, e_{3,k_3})$ and $P(E_6 | h_i, e_{1,k_1})$ are selected from the extreme mass functions of the credal set $K(E_j | h_i)$, $K(E_3 | h_i, e_{1,k_1})$, $K(E_4 | h_i, e_{3,k_3})$ and $K(E_6 | h_i, e_{1,k_1})$ to optimize the objective functions. 
 
In (\ref{eq:CFRInference}), $P(e_{j,k_j} | h_2)$ is the occurrence probability of the condition $e_{j, k_j}$ given an OTL failure is observed.
Considering the deficiency of the failure samples, the probability should be estimated using IDM.
On the other hand, $P(e_{j,k_j} | h_1)$ is the occurrence probability of the condition given no OTL failure happens.
In this case, abundant samples are available since no failure is a large probability event.
And this probability can be approximated using the average occurrence probability of the condition to simplify the calculation, e.g., $P(e_{1,3} | h_1)\approx P(e_{1,3})$ can be approximated by the statistical probability of high temperature in the relevant region.
The conditional probability $P(e_{3,k3} | h_i, e_{1,k1})$, $P(e_{4,k4} | h_i, e_{3,k3})$ and $P(e_{6,k6} | h_i, e_{1,k1})$ can be obtained in the same way.
Since the precise probability is a special case of the imprecise probability, the CFR can still be estimated by using Eq. (\ref{eq:CFRInference}) under this circumstance.

\section{Case Studies}

The proposed approach is tested by estimating the CFRs of two LGJ-300 transmission lines located in the same region.
One of the transmission lines, which is denoted as $\text{TL}_1$, has been in operation for 5 years.
Its recent three-year average failure rate is 0.00027 times per hour.
As a result, the parameter $P(h_1)$ and $P(h_2)$ of $\text{TL}_1$ are set to 0.99973 and 0.00027, respectively.
The other transmission line, denoted as $\text{TL}_2$, has been in operation for 10 years, and its recent three-year average failure rate is 0.00042 times per hour.
Therefore, the parameter $P(h_1)$ and $P(h_2)$ of $\text{TL}_2$ are set to 0.99958 and 0.00042, respectively.

A total of 40 failure samples are collected from the two transmission lines during their operation. The failure times under various operational conditions are listed in Table \ref{Table_FailureInformation}.

\begin{table}[!htb]\scriptsize
	\renewcommand{\arraystretch}{1.4}
	\caption{Failure Times under Various Operational Conditions}
	\vspace*{-6pt}
	\label{Table_FailureInformation}
	\newcommand{\tabincell}[2]{\begin{tabular}{@{}#1@{}}#2\end{tabular}}
	\centering
	\begin{tabular}{l r|l r}
		\hline
		Condition                                            & Times                            & Condition                                        & Times                              \\
	    \hline
		$T\le$4\textcelsius                                  &  $12$                            & Lightning, Rain                                  & $16$                               \\
		4\textcelsius$<T\le$26\textcelsius                   &  $7$                             & Lightning, No Rain                               & $2$                                \\
		$T>$26\textcelsius                                   &  $21$                            & No Lightning, Rain                               & $10$                               \\
		$W\le$12km/h                                         &  $8$                             & No Lightning, No Rain                            & $12$                               \\
		12km/h$<W\le$40km/h                                  &  $4$                             & $L\le80\%$                                       & $22$                               \\
		$T>$40km/h                                           &  $28$                            & $L>80\%$                                         & $18$                               \\
		Rain, $T\le$4\textcelsius                            &  $5$                             & Snow/Ice, $T\le$4\textcelsius                    & $8$                                \\
		Rain, 4\textcelsius$<T\le$26\textcelsius             &  $5$                             & Snow/Ice, 4\textcelsius$<T\le$26\textcelsius     & $0$                                \\		
	    Rain, $T>$26\textcelsius                             &  $16$                            & Snow/Ice, $T>$26\textcelsius                     & $0$                                \\	
		No Rain, $T\le$4\textcelsius                         &  $7$                             & No Snow/Ice, $T\le$4\textcelsius                 & $4$                                \\
		No Rain, 4\textcelsius$<T\le$26\textcelsius          &  $2$                             & No Snow/Ice, 4\textcelsius$<T\le$26\textcelsius  & $7$                                \\		
		No Rain, $T>$26\textcelsius                          &  $5$                             & No Snow/Ice, $T>$26\textcelsius                  & $21$                               \\
	    \hline
	\end{tabular}
\end{table}

\vspace*{-10pt}
\subsection{CFR Estimation by Dirichlet Model and Bayesian Network}

Deterministic CFR estimation can be performed by using the classical Bayesian network expressed by Eq. (\ref{eq:BNI}).
In the equation, the conditional mass functions are estimated according to Eq. (\ref{eq:Dirichlet_result}) with data listed in Table \ref{Table_FailureInformation}.
The selected prior weights for Eq. (\ref{eq:Dirichlet_result}) are listed in Table \ref{Table_PriorWeights}. 

\begin{table}[!htb]\scriptsize
	\renewcommand{\arraystretch}{1.4}
	\caption{Prior Weights for Conditional Mass Function Estimation}
	\vspace*{-6pt}
	\label{Table_PriorWeights}
	\newcommand{\tabincell}[2]{\begin{tabular}{@{}#1@{}}#2\end{tabular}}
	\centering
	\begin{tabular}{l r|l r}
		\hline
		Condition                                            & PW                                & Condition                                       & PW                                 \\
		\hline
		$T\le$4\textcelsius                                  &  $0.3$                            & Lightning, Rain                                 & $0.7$                              \\
		4\textcelsius$<T\le$26\textcelsius                   &  $0.1$                            & Lightning, No Rain                              & $0.1$                              \\
		$T>$26\textcelsius                                   &  $0.6$                            & No Lightning, Rain                              & $0.3$                              \\
		$W\le$12km/h                                         &  $0.2$                            & No Lightning, No Rain                           & $0.9$                              \\
		12km/h$<W\le$40km/h                                  &  $0.1$                            & $L\le80\%$                                      & $0.4$                              \\
		$T>$40km/h                                           &  $0.7$                            & $L>80\%$                                        & $0.6$                              \\
		Rain, $T\le$4\textcelsius                            &  $0.5$                            & Snow/Ice, $T\le$4\textcelsius                   & $0.7$                              \\
		Rain, 4\textcelsius$<T\le$26\textcelsius             &  $0.7$                            & Snow/Ice, 4\textcelsius$<T\le$26\textcelsius    & $0$                                \\		
		Rain, $T>$26\textcelsius                             &  $0.7$                            & Snow/Ice, $T>$26\textcelsius                    & $0$                                \\	
		No Rain, $T\le$4\textcelsius                         &  $0.5$                            & No Snow/Ice, $T\le$4\textcelsius                & $0.3$                              \\
		No Rain, 4\textcelsius$<T\le$26\textcelsius          &  $0.3$                            & No Snow/Ice, 4\textcelsius$<T\le$26\textcelsius & $1$                                \\		
		No Rain, $T>$26\textcelsius                          &  $0.3$                            & No Snow/Ice, $T>$26\textcelsius                 & $1$                                \\
		\hline
	\end{tabular}
\end{table}

On the basis of the information provided by Table \ref{Table_FailureInformation} and \ref{Table_PriorWeights}, the conditional mass function of hourly average temperature given an OTL failure is estimated, and the result is shown in Table \ref{Table_CMFofTemperature}.
In the table, the conditional mass function under no failure situation is approximated by the long-term statistical distribution of local temperature.

\begin{table}[!htb]
\renewcommand{\arraystretch}{1.2}
\caption{Conditional Mass Functions of Hourly Average Temperature}
\vspace*{-6pt}
\label{Table_CMFofTemperature}
\centering
\begin{tabular}{c c c c}
\hline
$h_i$ & $P(e_{1,1}|h_i)$ & $P(e_{1,2}|h_i)$ & $P(e_{1,3}|h_i)$\\
\hline
$h_1$ & 0.28             & 0.49             & 0.23            \\
$h_2$ & 0.30             & 0.17             & 0.53            \\
\hline
\end{tabular}
\end{table}

It can be observed from the table that when the transmission line $\text{TL}_1$ or $\text{TL}_2$ experiences a failure, the ambient temperature has a large probability of being higher than 26\textcelsius.

The conditional mass functions of other operational conditions can be obtained in the same way, and the results are listed in Table \ref{Table_CMFofWind}, \ref{Table_CMFofRain}, \ref{Table_CMFofLightning}, \ref{Table_CMFofLoad} and \ref{Table_CMFofSnow}.

\begin{table}[!htb]
	\renewcommand{\arraystretch}{1.2}
	\caption{Conditional Mass Functions of Hourly Average Wind Speed}
	\vspace*{-6pt}
	\label{Table_CMFofWind}
	\centering
	\begin{tabular}{c c c c}
		\hline
		$h_i$ & $P(e_{2,1}|h_i)$ & $P(e_{2,2}|h_i)$ & $P(e_{2,3}|h_i)$\\
		\hline
		$h_1$ & 0.40             & 0.42             & 0.18            \\
		$h_2$ & 0.20             & 0.10             & 0.70            \\
		\hline
	\end{tabular}
\end{table}

\begin{table}[!htb]
	\renewcommand{\arraystretch}{1.2}
	\caption{Conditional Mass Functions of Rain}
	\vspace*{-6pt}
	\label{Table_CMFofRain}
	\centering
	\begin{tabular}{c c c c}
		\hline
		$h_i$ & $e_{1,k_1}$      & $P(e_{3,1}|h_i,e_{1,k_1})$ & $P(e_{3,2}|h_i,e_{1,k_1})$\\
		\hline
		$h_1$ & $e_{1,1}$        & 0.04                       & 0.96            \\
		$h_1$ & $e_{1,2}$        & 0.12                       & 0.88            \\
		$h_1$ & $e_{1,3}$        & 0.21                       & 0.79            \\
		$h_2$ & $e_{1,1}$        & 0.40                       & 0.60            \\
		$h_2$ & $e_{1,2}$        & 0.69                       & 0.31            \\
		$h_2$ & $e_{1,3}$        & 0.75                       & 0.25            \\
		\hline
	\end{tabular}
\end{table}

\begin{table}[!htb]
	\renewcommand{\arraystretch}{1.2}
	\caption{Conditional Mass Functions of Lightning}
	\vspace*{-6pt}
	\label{Table_CMFofLightning}
	\centering
	\begin{tabular}{c c c c}
		\hline
		$h_i$ & $e_{3,k_3}$      & $P(e_{4,1}|h_i,e_{3,k_3})$ & $P(e_{4,2}|h_i,e_{3,k_3})$\\
		\hline
		$h_1$ & $e_{3,1}$        & 0.33                       & 0.67            \\
		$h_2$ & $e_{3,1}$        & 0.63                       & 0.37            \\
		$h_1$ & $e_{3,2}$        & 0.05                       & 0.95            \\
		$h_2$ & $e_{3,2}$        & 0.11                       & 0.89            \\
		\hline
	\end{tabular}
\end{table}

\begin{table}[!htb]
	\renewcommand{\arraystretch}{1.2}
	\caption{Conditional Mass Functions of Hourly Average Loading Rate}
	\vspace*{-6pt}
	\label{Table_CMFofLoad}
	\centering
	\begin{tabular}{c c c}
		\hline
		$h_i$ & $P(e_{5,1}|h_i)$ & $P(e_{5,2}|h_i)$ \\
		\hline
		$h_1$ & 0.67             & 0.33             \\
		$h_2$ & 0.55             & 0.45             \\
		\hline
	\end{tabular}
\end{table}

\begin{table}[!htb]
	\renewcommand{\arraystretch}{1.2}
	\caption{Conditional Mass Functions of Snow/Ice}
	\vspace*{-6pt}
	\label{Table_CMFofSnow}
	\centering
	\begin{tabular}{c c c c}
		\hline
		$h_i$ & $e_{1,k_1}$      & $P(e_{6,1}|h_i,e_{1,k_1})$ & $P(e_{6,2}|h_i,e_{1,k_1})$\\
		\hline
		$h_1$ & $e_{1,1}$        & 0.14                       & 0.86            \\
		$h_1$ & $e_{1,2}$        & 0.01                       & 0.99            \\
		$h_1$ & $e_{1,3}$        & 0                          & 1               \\
		$h_2$ & $e_{1,1}$        & 0.65                       & 0.35            \\
		$h_2$ & $e_{1,2}$        & 0                          & 1               \\
		$h_2$ & $e_{1,3}$        & 0                          & 1               \\

		\hline
	\end{tabular}
\end{table}

With respect to the aforementioned statistical results, the deterministic CFRs of the transmission lines can be estimated according to Eq. (\ref{eq:BNI}).
To illustrate this estimation approach, three scenarios with different operational conditions are tested.

\textbf{Scenario I:} The transmission line $\text{TL}_1$ is operating normally now. Its average loading rate will be 95\% in the next hour. At the same time, according to the short-term weather forecast, a thunderstorm will come in the next hour. During the storm, the average air temperature will be 30\textcelsius \ and the wind speed will be 48 km/h. Find the CFR of $\text{TL}_1$ for the coming hour.

\textbf{Scenario II:} The transmission line $\text{TL}_1$ is operating normally now. In the next hour the temperature will be 15\textcelsius, the wind speed will be 10 km/h and the loading rate of the transmission line will be less than 45\%. Find the CFR of $\text{TL}_1$ for the coming hour.

\textbf{Scenario III:} The transmission line $\text{TL}_2$ is operating normally now. All the operational conditions are the same as Scenario I. Estimate the CFR of $\text{TL}_2$ under this scenario.

The CFRs estimated by the Bayesian network are shown in Table \ref{Table_ResultofBN}.
It can be observed from the estimation results that the CFR of $\text{TL}_1$ under Scenario I is much higher than that under Scenario II.
This is because all the operational conditions of Scenario I are adverse to power transmission compared with that of Scenario II.
On the other hand, it can be found from the results of Scenario I and Scenario III that the CFR of $\text{TL}_2$ is obviously higher than that of $\text{TL}_1$ under the same weather and loading conditions.
$\text{TL}_2$ has a higher CFR because it is much older than TL1.

\begin{table}[!htb]
	\renewcommand{\arraystretch}{1.2}
	\caption{Estimation Results of the Bayesian Network}
	\vspace*{-6pt}
	\label{Table_ResultofBN}
	\centering
	\begin{tabular}{c c c}
		\hline
		Scenario I & Scenario II & Scenario III    \\
		\hline
		2.19E$-$2   & 1.16E$-$5    & 3.37E$-$2     \\
		\hline
	\end{tabular}
\end{table}

The estimation results of the deterministic approach may be influenced by the prior weights dramatically.
To illustrate this, a group of new prior weights are applied for the estimation, which are listed in Table \ref{Table_PriorWeights_1}.
The corresponding CFR estimation results are shown in Table \ref{Table_ResultofBN_1}.

\begin{table}[!htb]\scriptsize
	\renewcommand{\arraystretch}{1.2}
	\caption{Alternative Prior Weights for the Mass Function Estimation}
	\vspace*{-6pt}
	\label{Table_PriorWeights_1}
	\newcommand{\tabincell}[2]{\begin{tabular}{@{}#1@{}}#2\end{tabular}}
	\centering
	\begin{tabular}{l r|l r}
		\hline
		Condition                                            & PW                                & Condition                                       & PW                                 \\
		\hline
		$T\le$4\textcelsius                                  &  $0.4$                            & Lightning, Rain                                 & $0.5$                              \\
		4\textcelsius$<T\le$26\textcelsius                   &  $0.3$                            & Lightning, No Rain                              & $0.3$                              \\
		$T>$26\textcelsius                                   &  $0.3$                            & No Lightning, Rain                              & $0.5$                              \\
		$W\le$12km/h                                         &  $0.2$                            & No Lightning, No Rain                           & $0.7$                              \\
		12km/h$<W\le$40km/h                                  &  $0.4$                            & $L\le80\%$                                      & $0.8$                              \\
		$T>$40km/h                                           &  $0.4$                            & $L>80\%$                                        & $0.2$                              \\
		Rain, $T\le$4\textcelsius                            &  $0.2$                            & Snow/Ice, $T\le$4\textcelsius                   & $0.5$                              \\
		Rain, 4\textcelsius$<T\le$26\textcelsius             &  $0.5$                            & Snow/Ice, 4\textcelsius$<T\le$26\textcelsius    & $0$                                \\		
		Rain, $T>$26\textcelsius                             &  $0.5$                            & Snow/Ice, $T>$26\textcelsius                    & $0$                                \\	
		No Rain, $T\le$4\textcelsius                         &  $0.8$                            & No Snow/Ice, $T\le$4\textcelsius                & $0.5$                              \\
		No Rain, 4\textcelsius$<T\le$26\textcelsius          &  $0.5$                            & No Snow/Ice, 4\textcelsius$<T\le$26\textcelsius & $1$                                \\		
		No Rain, $T>$26\textcelsius                          &  $0.5$                            & No Snow/Ice, $T>$26\textcelsius                 & $1$                                \\
		\hline
	\end{tabular}
\end{table}

\begin{table}[!htb]
	\renewcommand{\arraystretch}{1.2}
	\caption{Estimation Results of the Bayesian Network}
	\vspace*{-6pt}
	\label{Table_ResultofBN_1}
	\centering
	\begin{tabular}{c c c}
		\hline
		Scenario I & Scenario II & Scenario III \\
		\hline
		2.05E$-$2   & 1.30E$-$5    & 3.15E$-$2     \\
		\hline
	\end{tabular}
\end{table}

By comparing the estimation results shown in Table \ref{Table_ResultofBN} and \ref{Table_ResultofBN_1}, it can be found that the estimated CFRs are different with different prior weights, which illustrates the significant influence of the prior weights on the CFR estimation results.

\subsection{CFR Estimation by IDM and Credal Network}

Using the historical observations shown in Table \ref{Table_FailureInformation}, the imprecise conditional mass functions given OTL failures can be obtained by using IDM according to Eq. (\ref{eq:IDM_result}), and the results are listed in Table \ref{Table_ICMFofTemperature}, \ref{Table_ICMFofWind}, \ref{Table_ICMFofRain}, \ref{Table_ICMFofLightning}, \ref{Table_ICMFofLoad} and \ref{Table_ICMFofSnow}.

\begin{table}[!htb]	
	\renewcommand{\arraystretch}{1.2}
	\caption{Imprecise Conditional Mass Function of Temperature}
	\vspace*{-6pt}
	\label{Table_ICMFofTemperature}
	\centering
	\begin{tabular}{c c c c}
		\hline
		$h_i$ & $P(e_{1,1}|h_i)$ & $P(e_{1,2}|h_i)$ & $P(e_{1,3}|h_i)$\\
		\hline
		$h_2$ & [0.29, 0.32]     & [0.17, 0.20]      & [0.51, 0.54]            \\
		\hline
	\end{tabular}
\end{table}

\begin{table}[!htb]
	\renewcommand{\arraystretch}{1.2}
	\caption{Imprecise Conditional Mass Function of Wind Speed}
	\vspace*{-6pt}
	\label{Table_ICMFofWind}
	\centering
	\begin{tabular}{c c c c}
		\hline
		$h_i$ & $P(e_{2,1}|h_i)$ & $P(e_{2,2}|h_i)$ & $P(e_{2,3}|h_i)$\\
		\hline
		$h_2$ & [0.19, 0.22]     & [0.10, 0.13]      & [0.68, 0.71]            \\
		\hline
	\end{tabular}
\end{table}

\begin{table}[!htb]
	\renewcommand{\arraystretch}{1.2}
	\caption{Imprecise Conditional Mass Function of Rain}
	\vspace*{-6pt}
	\label{Table_ICMFofRain}
	\centering
	\begin{tabular}{c c c c}
		\hline
		$h_i$ & $e_{1,k_1}$      & $P(e_{3,1}|h_i,e_{1,k_1})$ & $P(e_{3,2}|h_i,e_{1,k_1})$\\
		\hline
		$h_2$ & $e_{1,1}$        & [0.38, 0.46]               & [0.54, 0.62]            \\
		$h_2$ & $e_{1,2}$        & [0.63, 0.75]               & [0.25, 0.37]            \\
		$h_2$ & $e_{1,3}$        & [0.73, 0.77]               & [0.23, 0.27]            \\
		\hline
	\end{tabular}
\end{table}

\begin{table}[!htb]
	\renewcommand{\arraystretch}{1.2}
	\caption{Imprecise Conditional Mass Function of Lightning}
	\vspace*{-6pt}
	\label{Table_ICMFofLightning}
	\centering
	\begin{tabular}{c c c c}
		\hline
		$h_i$ & $e_{3,k_3}$      & $P(e_{4,1}|h_i,e_{3,k_3})$ & $P(e_{4,2}|h_i,e_{3,k_3})$\\
		\hline
		$h_2$ & $e_{3,1}$        & [0.59, 0.63]               & [0.37, 0.41]            \\
		$h_2$ & $e_{3,2}$        & [0.08, 0.15]               & [0.85, 0.92]            \\
		\hline
	\end{tabular}
\end{table}

\begin{table}[!htb]
	\renewcommand{\arraystretch}{1.2}
	\caption{Imprecise Conditional Mass Function of Hourly Average Loading Rate}
	\vspace*{-6pt}
	\label{Table_ICMFofLoad}
	\centering
	\begin{tabular}{c c c}
		\hline
		$h_i$ & $P(e_{5,1}|h_i)$ & $P(e_{5,2}|h_i)$ \\
		\hline
		$h_1$ & [0.54, 0.56]     & [0.44, 0.46]     \\
		\hline
	\end{tabular}
\end{table}

\begin{table}[!htb]
	\renewcommand{\arraystretch}{1.2}
	\caption{Imprecise Conditional Mass Function of Snow/Ice}
	\vspace*{-6pt}
	\label{Table_ICMFofSnow}
	\centering
	\begin{tabular}{c c c c}
		\hline
		$h_i$ & $e_{6,k_1}$      & $P(e_{6,1}|h_i,e_{1,k_1})$ & $P(e_{6,2}|h_i,e_{1,k_1})$\\
		\hline
		$h_2$ & $e_{1,1}$        & [0.62, 0.69]               & [0.31, 0.38]            \\
		$h_2$ & $e_{1,2}$        & [0, 0.12]                  & [0.88, 1]            \\
		$h_2$ & $e_{1,3}$        & [0, 0.04]                  & [0.96, 1]            \\
		\hline
	\end{tabular}
\end{table}

It is observed from the tables that the conditional probabilities estimated by the deterministic Dirichlet model are all included in the probability intervals estimated by the IDM, which illustrates the rationality of the IDM estimation results.
The widths of the probability intervals estimated by the IDM quantitatively reflect the uncertainty existed in the probability estimation results.

Respecting the estimation results of IDM, the imprecise CFRs corresponding to the three scenarios can be calculated according to Eq. (\ref{eq:CFRInference}), and the results are listed in Table \ref{Table_ResultofCN}. 

\begin{table}[!htb]
	\renewcommand{\arraystretch}{1.2}
	\caption{Estimation Results of the Credal Network}
	\vspace*{-6pt}
	\label{Table_ResultofCN}
	\centering
	\begin{tabular}{c c c}
		\hline
		Scenario I & Scenario II & Scenario III \\
		\hline
		[1.85E$-$2, 2.38E$-$2]   &[8.88E$-$6, 1.94E$-$5]    & [2.85E$-$2, 3.66E$-$2]     \\
		\hline
	\end{tabular}
\end{table}

It can be found that the estimation results of the Bayesian network with different prior weights are all included in the probability intervals estimated by the credal network.
In fact, this phenomenon is universal, which verifies that the proposed approach can eliminate the subjectivity caused by the prior weight assignment.
Furthermore, it can be found by comparing the estimation results corresponding to different scenarios that the effects of the weather, loading and aging conditions on the CFRs can also be reflected in the estimation results of the credal network.

Sometimes, the occurrence of the operational conditions is uncertain.
For instance, it is difficult to exactly predict the air temperature in the coming hour.
The proposed approach can handle such uncertainty conveniently by using the law of total probability, as illustrated by the following example.

\textbf{Scenario IV:} The transmission line $\text{TL}_2$ is operating normally now. All the operational conditions are the same as in Scenario I except that the wind speed has a 70\% probability of being faster than 40 km/h and a 30\% probability of ranging from 12 km/h to 40 km/h.

Under this scenario, the CFR is estimated to be within [$2.05\times10^{-2}$, $2.65\times10^{-2}$].
Comparing this result with the imprecise CFR estimation result of Scenario III, it can be found that the CFR is lower under Scenario IV.
This is because the wind speed has a considerable probability of being slow in Scenario IV, which is beneficial for the power transmission.

\subsection{Comparison with Other Probability Interval Estimation Approaches}

\textcolor{black}{
A simulation system is set up to compare the performance of the proposed approach and two existing approaches, i.e., the central limit theorem based approach \cite{zhou2006modeling} and the Chi-square distribution based approach \cite{li2008fuzzy}.
To simplify the analysis, the operational conditions are categorized into two groups, i.e., the normal operational condition and the adverse operational condition, and the occurrence probabilities of these two operational conditions are 0.7 and 0.3, respectively.
Assume the failure rates of the interested OTL under normal and adverse operational conditions are 0.0001 and 0.005, respectively.
The simulation system is virtually operated for 10000 hours, and the collected failure and operational condition data are used to estimate the CFR of the OTL.
}

\begin{figure}[!htb]
	\centering
	\includegraphics[width=3.5in]{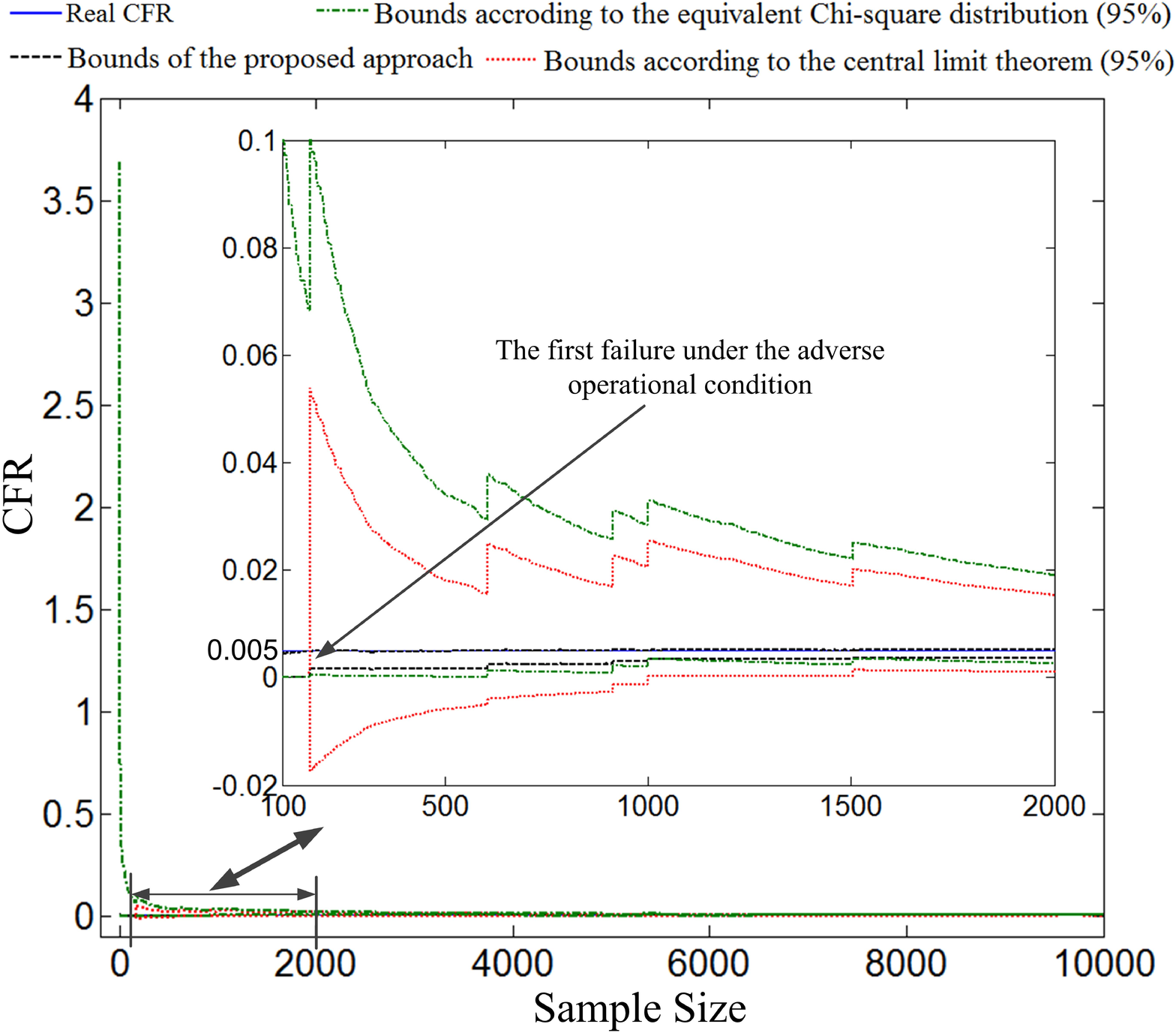}
	\caption{CFR intervals under the adverse operational condition.}
	\label{CFRIntervalCompare}
\end{figure}

\textcolor{black}{
	The CFR interval under the adverse operational condition is estimated by using the aforementioned approaches, and the results are shown in Fig. \ref{CFRIntervalCompare}.}

\textcolor{black}{
	From the figure, the following facts can be observed.
	\begin{enumerate}
		\item The CFR intervals estimated by all the approaches can converge to the real CFR as data accumulate. Meanwhile, the interval of the proposed approach converges much faster than those of the existing approaches.
		\item When the sample size is small, the CFR interval estimated by the Chi-square distribution based approach is very large. The upper bound of the estimated interval is even greater than 1, mainly because the equivalent Chi-square distribution is unbounded on the right side.
		\item The central limit theorem based approach cannot perform the interval estimation until the occurrence of the first failure. Meanwhile, the lower bound of the estimated interval may be less than 0 because the converted normal distribution is unbounded.
	\end{enumerate}
}

\textcolor{black}{
The observations illustrate that the proposed approach can obtain more reasonable and more accurate CFR intervals than the existing approaches.
}

\section{Conclusion}

Because of the deficiency of the OTL failure samples, the CFR estimation result may be unreliable in practice.
Under this circumstance, a novel imprecise probabilistic approach for the CFR estimation, based on IDM and the credal network, is proposed.
In the approach, IDM is adopted to estimate the imprecise probabilistic dependency relations between the OTL failure and the operational conditions, and the credal network is established to integrate the IDM estimation results and infer the imprecise CFR.
The proposed approach is tested by estimating the CFRs of two transmission lines located in the same region.
The test results illustrate that:
a) the proposed approach can quantitatively evaluate the uncertainty of the estimated CFR by using the interval probability, b) the influences of the operational conditions on the CFR can be properly reflected in the estimation result,
and c) the uncertainty of the operational conditions can also be handled by using the proposed approach.
\textcolor{black}{
Moreover, a simulation system is set up to demonstrate the advantages of the proposed approach over the existing approaches, i.e., the central limit theorem based approach and the Chi-square distribution based approach.
The test results indicate that the CFR interval estimated by the proposed approach is much tighter and more reasonable than those obtained by the existing approaches.
}

% or


\appendices
\section{The Credible Interval}
\textcolor{black}{
The credible interval $D=[a,b]$ is the interval that ensures the posterior lower probability $\underline{\text{Pr}}\{P_m\in D\}$ reaches a specified credibility $\gamma$, e.g., $\gamma=0.95$, where $P_m$ is the probability corresponding to the $m$th outcome of the multinomial variable.}

\textcolor{black}{ 
The bounds of $D$ can be obtained by
\begin{align}\label{eq:Appendix}
	\begin{cases}
	a=0, \ \ b=G^{-1} \big( \frac{1+\gamma}{2} \big), \ \ & n_m=0, \\
	a=H^{-1} \big( \frac{1-\gamma}{2} \big),\ \ b=G^{-1} \big( \frac{1+\gamma}{2} \big), \ \  &0<n_m<n, \\
	a=H^{-1} \big( \frac{1-\gamma}{2} \big),\ \ b=1,\ \ &n_m=n,	
	\end{cases}
\end{align}
where 
$H$ is the cumulative distribution function (CDF) of beta distribution $B(n_m,s+n-n_m)$,
$G$ is the CDF of beta distribution $B(s+n_m, n-n_m)$,
$n$ is the sample size,
$n_m$ is the number of times that the $m$th outcome is observed,
and $s$ is the equivalent sample size that is set to 1 \cite{walley1996inferences}.
}
\section{Comparison of the Probability Interval Counting Approaches}

\begin{figure}[!htb]
	\centering
	\includegraphics[width=3.2in]{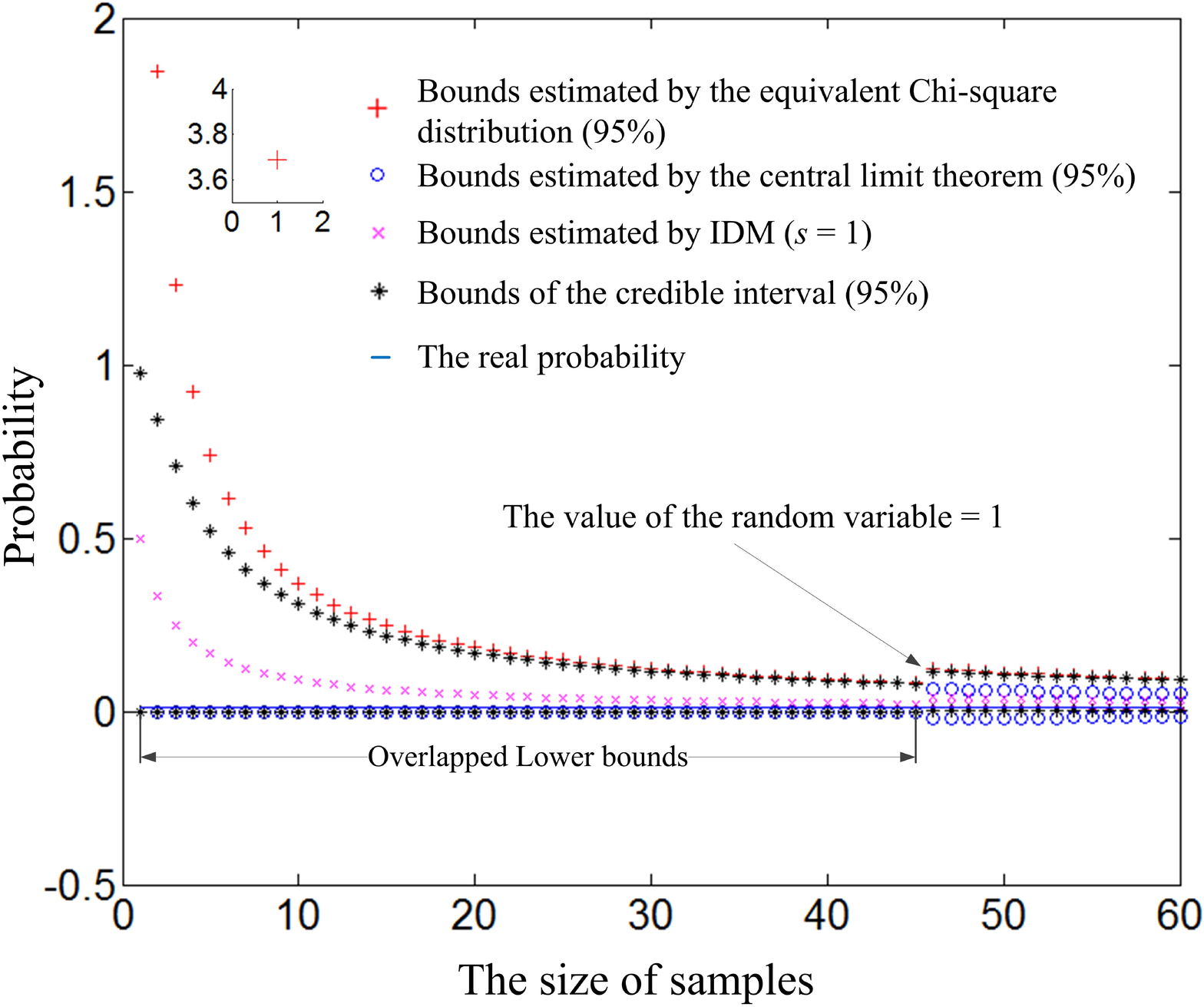}
	\caption{Probability intervals estimated by different methods.}
	\label{AppendixB}
\end{figure}

\textcolor{black}{
Probability intervals counted from available samples are the basis of the proposed approach.
The performance of several counting approaches, i.e., IDM, the credible interval estimation approach, the central limit theorem based approach \cite{zhou2006modeling}, and the Chi-square distribution based approach \cite{li2008fuzzy}, is tested.
The samples are randomly selected from Poisson distribution $\text{Pois}(0.01)$, which means for each sample the corresponding two-state random variable has a 1\% chance to be 1 and a 99\% chance to be 0.
The test results are shown in Fig. \ref{AppendixB}.
}

\textcolor{black}{
From the results, the following facts can be observed.
}
\textcolor{black}{
\begin{enumerate}
	\item IDM can obtain the tightest probability interval that also converges much faster than the other approaches. Meanwhile, when the sample size is smaller than 25, the 95\% credible interval obtained by the approach in Appendix A is obviously tighter than the 95\% confidence interval estimated by the Chi-square distribution based approach.
	\item The upper bound of the probability interval estimated by the Chi-square distribution based approach may be greater than 1 (see the confidence intervals corresponding to the first three samples), which is obviously unreasonable.
	\item The central limit theorem based approach cannot obtain a meaningful probability interval until all the states of the two-state variable occur at least once. Moreover, the estimated lower bound may be less than 0 because the converted normal distribution is unbounded.
\end{enumerate}
}

\textcolor{black}{
The observations illustrate that IDM and the corresponding credible interval estimation approach can provide more reasonable estimation results compared with the existing approaches.}

\textcolor{black}{
In this paper, IDM is used to estimate the uncertain probabilistic dependency relations between the OTL failure and the operational conditions.
Meanwhile, the credible interval estimation approach mentioned in Appendix A is recommended as an alternative if a quantitative confidence index is required.}

\section{Effects of the Equivalent Sample Size}

\textcolor{black}{
The equivalent sample size $s$ can influence the estimation result of IDM. In fact, $s$ determines how quickly the probability interval will converge as data accumulate. Smaller values of $s$ produce faster convergence and stronger conclusions, whereas larger values of $s$ produce more cautious inferences. More specifically, $s$ indicates the number of observations needed to reduce the length of the imprecision interval to half of its initial value (according to Eq. (\ref{eq:IDM_result}), the length of the imprecision interval is $\Delta P_m = \overline{P}_m - \underline{P}_m = s/(n+s)$, which will decrease from $1$ to $1/2$ when $n$ increases from $0$ to $s$).
}

\begin{figure}[!htb]
	\centering
	\includegraphics[width=3.2in]{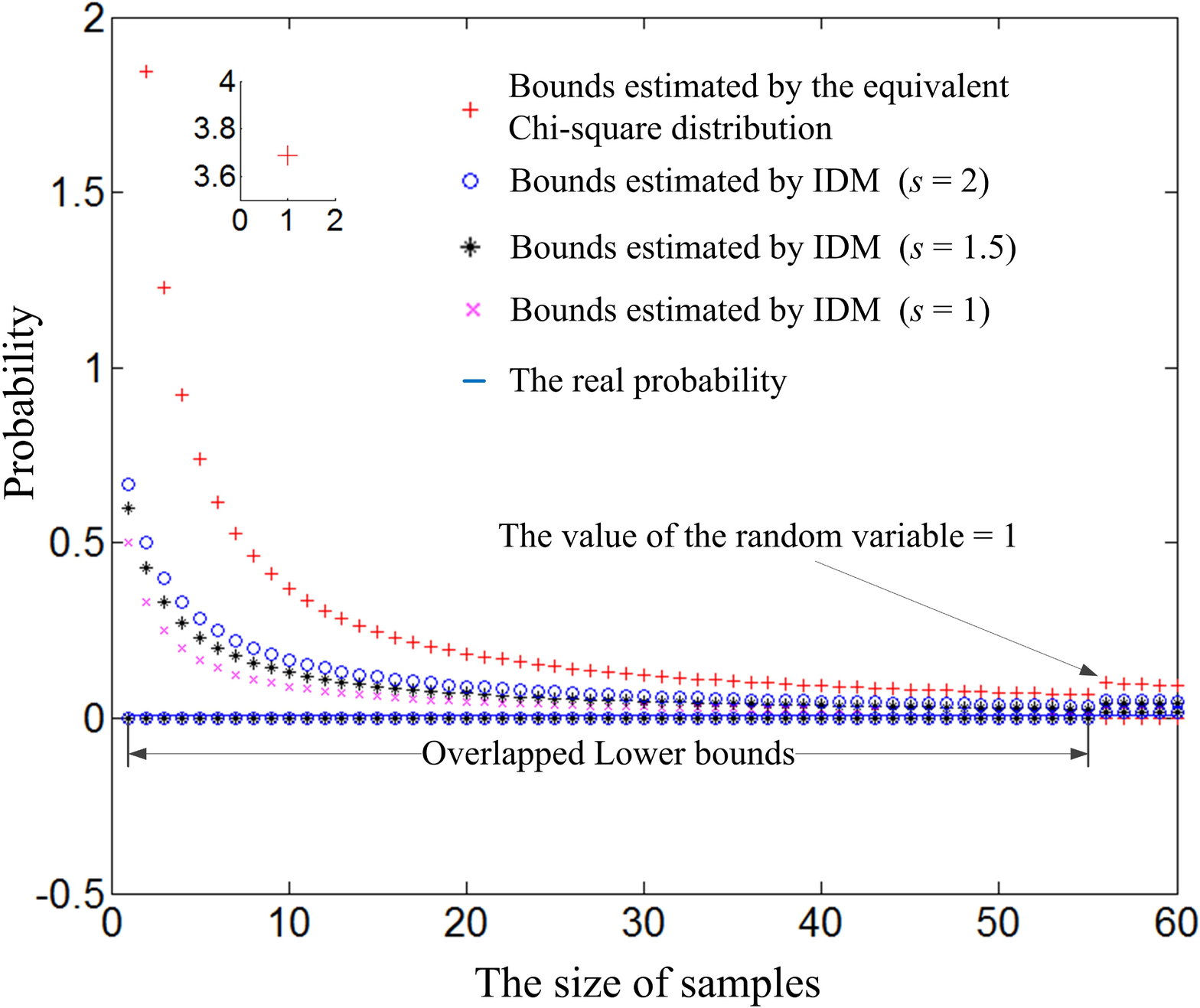}
	\caption{Probability intervals estimated with different equivalent sample sizes.}
	\label{AppendixC}
\end{figure}

\textcolor{black}{
Therefore, the assignment of $s$ reflects the cautiousness of the estimation.
To obtain objective estimation results, as suggested by Walley in \cite{walley1996inferences}, the value of $s$ should be sufficient large to encompass all reasonable probability intervals estimated by various objective Bayesian alternatives.
Up to now, several researches have led to convincing arguments that $s$ should be chosen from $[1,2]$, and the value of $s$ should be larger when the multinomial variable has a large number of states \cite{bernard2005introduction}.
In our model, since the multinomial variables only have 2 or 3 states (see Table I), the parameter $s$ is set to 1 as suggested.
Moreover, it should be emphasized that the influence of $s$ weakens quickly as the number of samples increases, which can be clearly seen from Fig. \ref{AppendixC}.
It can also be observed from the figure that all of the probability intervals estimated by IDM ($1\le s \le 2$) converge faster than that estimated by the Chi-square distribution based approach.
}

\ifCLASSOPTIONcaptionsoff
  \newpage
\fi

\footnotesize
\bibliographystyle{IEEEtran}
\bibliography{Reference}

\begin{thebibliography}{10}
\providecommand{\url}[1]{#1}
\csname url@samestyle\endcsname
\providecommand{\newblock}{\relax}
\providecommand{\bibinfo}[2]{#2}
\providecommand{\BIBentrySTDinterwordspacing}{\spaceskip=0pt\relax}
\providecommand{\BIBentryALTinterwordstretchfactor}{4}
\providecommand{\BIBentryALTinterwordspacing}{\spaceskip=\fontdimen2\font plus
\BIBentryALTinterwordstretchfactor\fontdimen3\font minus
  \fontdimen4\font\relax}
\providecommand{\BIBforeignlanguage}[2]{{%
\expandafter\ifx\csname l@#1\endcsname\relax
\typeout{** WARNING: IEEEtran.bst: No hyphenation pattern has been}%
\typeout{** loaded for the language `#1'. Using the pattern for}%
\typeout{** the default language instead.}%
\else
\language=\csname l@#1\endcsname
\fi
#2}}
\providecommand{\BIBdecl}{\relax}
\BIBdecl

\bibitem{allan2013reliability}
R.~Allan \emph{et~al.}, \emph{Reliability evaluation of power systems}.\hskip
  1em plus 0.5em minus 0.4em\relax Springer Science \& Business Media, 2013.

\bibitem{billinton1999teaching}
R.~Billinton and P.~Wang, ``Teaching distribution system reliability evaluation
  using {Monte Carlo} simulation,'' \emph{IEEE Trans. Power Syst.}, vol.~14,
  no.~2, pp. 397--403, Aug. 1999.

\bibitem{qi2013blackout}
J.~Qi, S.~Mei, and F.~Liu, ``Blackout model considering slow process,''
  \emph{IEEE Trans. Power Syst.}, vol.~28, no.~3, pp. 3274--3282, Aug. 2013.

\bibitem{koval2009assessment}
D.~Koval and A.~Chowdhury, ``Assessment of transmission-line common-mode,
  station-originated, and fault-type forced-outage rates,'' \emph{IEEE Trans.
  Ind. Appl.}, vol.~46, no.~1, pp. 313--318, Jan. 2010.

\bibitem{qi2015interaction}
J.~Qi, K.~Sun, and S.~Mei, ``An interaction model for simulation and mitigation
  of cascading failures,'' \emph{IEEE Trans. Power Syst.}, vol.~30, no.~2, pp.
  804--819, Jul. 2015.

\bibitem{qi2016estimating}
J.~Qi, W.~Ju, and K.~Sun, ``Estimating the propagation of interdependent
  cascading outages with multi-type branching processes,'' \emph{IEEE Trans.
  Power Syst.}, to be published.

\bibitem{williams2003weather}
C.~Williams, ``Weather normalization of power system reliability indices,'' in
  \emph{IEEE Power Eng. Soc. General Meeting}.\hskip 1em plus 0.5em minus
  0.4em\relax Tampa, FL, USA, Sep. 7-12, 2007, pp. 1--5.

\bibitem{ning2011time}
L.~Ning, W.~Wu, B.~Zhang, and P.~Zhang, ``A time-varying transformer outage
  model for on-line operational risk assessment,'' \emph{Int. J. Elec. Power},
  vol.~33, no.~3, pp. 600--607, Mar. 2011.

\bibitem{feng2008power}
Y.~Feng, W.~Wu, B.~Zhang, and W.~Li, ``Power system operation risk assessment
  using credibility theory,'' \emph{IEEE Trans. Power Syst.}, vol.~23, no.~3,
  pp. 1309--1318, Aug. 2008.

\bibitem{billinton2006application}
R.~Billinton and G.~Singh, ``Application of adverse and extreme adverse
  weather: modelling in transmission and distribution system reliability
  evaluation,'' \emph{Proc. Inst. Elect. Eng., Gen., Transm. Distrib.}, vol.
  153, no.~1, pp. 115--120, Jan. 2006.

\bibitem{billinton1991novel}
R.~Billinton and W.~Li, ``A novel method for incorporating weather effects in
  composite system adequacy evaluation,'' \emph{IEEE Trans. Power Syst.},
  vol.~6, no.~3, pp. 1154--1160, Aug. 1991.

\bibitem{zhou2006modeling}
Y.~Zhou, A.~Pahwa, and S.~S. Yang, ``Modeling weather-related failures of
  overhead distribution lines,'' \emph{IEEE Trans. Power Syst.}, vol.~21,
  no.~4, pp. 1683--1690, Nov. 2006.

\bibitem{li2008fuzzy}
W.~Li, J.~Zhou, and X.~Xiong, ``Fuzzy models of overhead power line
  weather-related outages,'' \emph{IEEE Trans. Power Syst.}, vol.~3, no.~23,
  pp. 1529--1531, Aug. 2008.

\bibitem{willis2000panel}
H.~L. Willis, ``Panel session on: aging t\&d infrastructures and customer
  service reliability,'' in \emph{IEEE Power Eng. Soc. Summer Meeting},
  vol.~3.\hskip 1em plus 0.5em minus 0.4em\relax Seattle, WA, USA, Jul. 16-20,
  2000, pp. 1494--1496.

\bibitem{carer2008weather}
P.~Carer and C.~Briend, ``Weather impact on components reliability: A model for
  mv electrical networks,'' in \emph{Proc. the 10th Int. Conf. PMAPS,}.\hskip
  1em plus 0.5em minus 0.4em\relax Rincon, Puerto Rico, USA, May 25-29, 2008,
  pp. 1--7.

\bibitem{bollen2000effects}
M.~Bollen, ``Effects of adverse weather and aging on power system
  reliability,'' \emph{IEEE Trans. Ind. Appl.}, vol.~37, no.~2, pp. 452--457,
  Mar. 2001.

\bibitem{ding2008fuzzy}
Y.~Ding, M.~J. Zuo, A.~Lisnianski, and Z.~Tian, ``Fuzzy multi-state systems:
  general definitions, and performance assessment,'' \emph{IEEE Trans.
  Reliab.}, vol.~57, no.~4, pp. 589--594, Dec. 2008.

\bibitem{ding2008fuzzy1}
Y.~Ding and A.~Lisnianski, ``Fuzzy universal generating functions for
  multi-state system reliability assessment,'' \emph{Fuzzy Set. Syst.}, vol.
  159, no.~3, pp. 307--324, Feb. 2008.

\bibitem{walpole1993probability}
R.~E. Walpole, R.~H. Myers, S.~L. Myers, and K.~Ye, \emph{Probability and
  statistics for engineers and scientists}.\hskip 1em plus 0.5em minus
  0.4em\relax Macmillan New York, 1993, vol.~5.

\bibitem{johnson2005univariate}
N.~L. Johnson, A.~W. Kemp, and S.~Kotz, \emph{Univariate discrete
  distributions}.\hskip 1em plus 0.5em minus 0.4em\relax John Wiley \& Sons,
  2005, vol. 444.

\bibitem{walley1991statistical}
P.~Walley, \emph{Statistical Reasoning with Imprecise Probabilities}.\hskip 1em
  plus 0.5em minus 0.4em\relax Chapman \& Hall, 1991.

\bibitem{coolen1997imprecise}
F.~Coolen, ``An imprecise {Dirichlet} model for {Bayesian} analysis of failure
  data including right-censored observations,'' \emph{Reliab. Eng. Syst.
  Safe.}, vol.~56, no.~1, pp. 61--68, Apr. 1997.

\bibitem{li2011interval}
C.~Li, X.~Chen, X.~Yi, and J.~Tao, ``Interval-valued reliability analysis of
  multi-state systems,'' \emph{IEEE Trans. Reliab.}, vol.~60, no.~1, pp.
  323--330, Mar. 2011.

\bibitem{finkelstein2008failure}
M.~Finkelstein, \emph{Failure Rate Modelling for Reliability and Risk}.\hskip
  1em plus 0.5em minus 0.4em\relax Springer, 2008.

\bibitem{bernard2005introduction}
J.~M. Bernard, ``An introduction to the imprecise {Dirichlet} model for
  multinomial data,'' \emph{Int. J. Approx. Reason.}, vol.~39, no.~2, pp.
  123--150, Jun. 2005.

\bibitem{masegosa2014imprecise}
A.~R. Masegosa and S.~Moral, ``Imprecise probability models for learning
  multinomial distributions from data. {Applications} to learning credal
  networks,'' \emph{Int. J. Approx. Reason.}, vol.~55, no.~7, pp. 1548--1569,
  Oct. 2014.

\bibitem{walley1996inferences}
P.~Walley, ``Inferences from multinomial data: learning about a bag of
  marbles,'' \emph{J. R. Stat. Soc. B}, vol.~58, no.~1, pp. 3--57, Jan. 1996.

\bibitem{antonucci2007credal}
A.~Antonucci, A.~Piatti, and M.~Zaffalon, ``Credal networks for operational
  risk measurement and management,'' in \emph{Knowledge-Based Intelligent
  Information and Engineering Systems}.\hskip 1em plus 0.5em minus 0.4em\relax
  Springer, 2007, pp. 604--611.

\bibitem{heckerman1998tutorial}
D.~Heckerman, \emph{A Tutorial on Learning with Bayesian Networks}.\hskip 1em
  plus 0.5em minus 0.4em\relax Springer, 1998.

\bibitem{cozman2000credal}
F.~G. Cozman, ``Credal networks,'' \emph{Artif. Intell.}, vol. 120, no.~2, pp.
  199--233, Jul. 2000.

\bibitem{cozman2005graphical}
{F. G. Cozman}, ``Graphical models for imprecise probabilities,'' \emph{Int. J.
  Approx. Reason.}, vol.~39, no.~2, pp. 167--184, Jun. 2005.

\bibitem{abellan2006measures}
J.~Abell{\'a}n and M.~G{\'o}mez, ``Measures of divergence on credal sets,''
  \emph{Fuzzy Set. Syst.}, vol. 157, no.~11, pp. 1514--1531, Jun. 2006.

\bibitem{corani2012bayesian}
G.~Corani, A.~Antonucci, and M.~Zaffalon, ``Bayesian networks with imprecise
  probabilities: Theory and application to classification,'' in \emph{Data
  Mining: Foundations and Intelligent Paradigms}.\hskip 1em plus 0.5em minus
  0.4em\relax Springer, 2012, pp. 49--93.

\bibitem{christiaanse1971reliability}
W.~Christiaanse, ``Reliability calculations including the effects of overloads
  and maintenance,'' \emph{IEEE Trans. Power Appa. \& Syst.}, no.~4, pp.
  1664--1677, Jul. 1971.

\end{thebibliography}

\end{document}